\documentclass[conference]{IEEEtran}

\usepackage{amsmath,amssymb,amsfonts}
\usepackage{algorithmic}
\usepackage{graphicx}
\usepackage[caption=false]{subfig}
\usepackage{textcomp}
\usepackage{xcolor}
\usepackage{balance}
\usepackage{cite}
\usepackage[colorlinks=true]{hyperref}
\usepackage{listings}
\def\BibTeX{{\rm B\kern-.05em{\sc i\kern-.025em b}\kern-.08em
    T\kern-.1667em\lower.7ex\hbox{E}\kern-.125emX}}

\newcommand{\reffig}[1]{Fig.~\ref{#1}}
\newcommand{\reftab}[1]{TABLE~\ref{#1}}
\newcommand{\refsec}[1]{Section~\ref{#1}}
\newcommand{\refequ}[1]{Equation~\ref{#1}}

\begin{document}

\title{mpiQulacs: A Distributed Quantum Computer Simulator for A64FX-based Cluster Systems}

\author{
    \IEEEauthorblockN{Satoshi Imamura, Masafumi Yamazaki, Takumi Honda, Akihiko Kasagi,\\ Akihiro Tabuchi, Hiroshi Nakao, Naoto Fukumoto, and Kohta Nakashima}
    \IEEEauthorblockA{\textit{ICT Systems Laboratory}\\
    \textit{Fujitsu LTD.}}
}

\maketitle
\thispagestyle{plain}
\pagestyle{plain}

\begin{abstract}
Quantum computer simulators running on classical computers are essential for developing real quantum computers and emerging quantum applications. In particular, state vector simulators, which store a full state vector in memory and update it in every quantum operation, are available to simulate an arbitrary form of quantum circuits, debug quantum applications, and validate future quantum computers. However, the time and space complexity grows exponentially with the number of qubits and easily exceeds the capability of a single machine.

Therefore, we develop a distributed state vector simulator, {\it mpiQulacs}, that is optimized for large-scale simulation on A64FX-based cluster systems. A64FX is an ARM-based CPU that is also equipped in the world's top Fugaku supercomputer. We evaluate weak and strong scaling of mpiQulacs with up to 36 qubits on a new 64-node A64FX-based cluster system named {\it Todoroki}. By comparing mpiQulacs with existing distributed state vector simulators, we show that mpiQulacs achieves the highest performance for large-scale simulation on tens of nodes while sustaining a nearly ideal scalability. Besides, we define a new metric, {\it quantum B/F ratio}, and use it to demonstrate that mpiQulacs running on Todoroki fits the requirements of distributed state vector simulation rather than the existing simulators running on general purpose CPU-based or GPU-based cluster systems.

\end{abstract}


\section{Introduction}
\label{sec:introduction}

Quantum computing is attracting a lot of attention both in industry and academia owing to the potential of exponential computing power. However, since it is still at the dawn, further innovation and development both in quantum hardware and software are necessary for the practical usage of quantum computing. Quantum computer simulators running on classical computers are essential for this purpose, because they can simulate the behaviors of future quantum computers and accelerate the development of emerging quantum applications.

{\it State vector} simulators are one of the representative quantum computer simulators, which store a full state vector in memory and update it in every quantum operation. They can be used to simulate an arbitrary form of quantum circuits, debug quantum applications, and validate real quantum computers. Although {\it tensor network} simulators have also been developed to simulate large-scale quantum circuits with up to 100 qubits \cite{Liu:2021cl}, they can only simulate circuits with low depth and do not effectively support intermediate measurement for quantum software debugging \cite{Wu:2019fu}. We aim to accelerate the development of emerging quantum applications and thus target state vector simulators in this work.

The major challenge of state vector simulators is time and space complexity which grows exponentially with the number of qubits of a given quantum circuit. When a state vector is represented in double precision, an $n$-qubit circuit requires $2^{n+4}$ bytes of memory space. For instance, a 36-qubit circuit requires 1~TiB of memory space. Although such a memory space can be allocated on a single high-end server having terabytes of memory, its computational power is not sufficient for the 36-qubit simulation in a practical time. On the other hand, high-performance computing devices such as GPUs are necessary for fast simulation, but the memory capacity of such devices is still limited to tens of gigabytes. Therefore, distributed state vector simulators have been developed for fast and large-scale simulation on cluster systems \cite{intelqs:2020,quest:2019,qiskitaer:2020,juqcsg:2021}.

In this work, we develop a distributed state vector simulator, {\it mpiQulacs}, that is based on {\it Qulacs} \cite{qulacs:2021} and optimized for A64FX-based cluster systems. Qulacs is one of the fastest state vector simulator running on a single machine and supports general quantum operations to meet popular demands in quantum computing research. We extend Qulacs with message passing interface (MPI) to support distributed simulation on multiple computing nodes. mpiQulacs is optimized to fully utilize the high memory bandwidth of an A64FX CPU. It is a 48-core ARM-based CPU having high-bandwidth memory (HBM2) and also used in the world's top Fugaku supercomputer. In addition, we implement a {\it fused-swap} gate in mpiQulacs based on the idea discussed in \cite{Raedt:2006ma} to minimize MPI communication for distributed simulation. 

We construct a new 64-node A64FX-based cluster system named {\it Todoroki} and use it to evaluate weak and strong scaling of mpiQulacs with up to 36 qubits. By comparing the results of mpiQulacs with those of existing distributed state vector simulators reported in corresponding papers, we show that mpiQulacs achieves the highest performance for large-scale simulation on tens of nodes while sustaining nearly ideal weak and strong scaling. Moreover, we define a novel metric, {\it quantum B/F ratio (QBF)}, that indicates the execution efficiency of state vector simulators running on cluster systems. The QBF evaluation demonstrates that mpiQulacs running on Todoroki fits the requirements of distributed state vector simulation rather than the existing simulators running on general purpose CPU-based or GPU-based cluster systems.

The rest of the paper is organized as follows. \refsec{sec:background} introduces the basis of state vector simulation and four existing distributed state vector simulators. \refsec{sec:mpiqulacs} explains the optimizations of mpiQulacs for the A64FX-based cluster system. \refsec{sec:evaluation} describes our experimental setup and shows the comparison results between mpiQulacs and the existing simulators. Finally, we conclude in \refsec{sec:conclusion}.
\section{Background}
\label{sec:background}

In this section, we first explain the basis of state vector simulation and then introduce four existing distributed state vector simulators.

\subsection{Basis of State Vector Simulation}

Quantum computing is based on a {\it qubit}, which is the minimum unit of quantum information. While a classical bit represents either of zero or one, a qubit can represent a superposition of both states simultaneously. A qubit state is expressed as 
\begin{equation}
  |\psi\rangle = a_{0}|0\rangle + a_{1}|1\rangle
\end{equation}
where $a_{0}$ and $a_{1}$ are complex probability amplitudes and $|a_{0}|^{2}+|a_{1}|^2=1$. $|0\rangle$ and $|1\rangle$ are two orthonormal basis states, and the probability of outcome $|0\rangle$ or $|1\rangle$ is $|a_{0}|^2$ or $|a_{1}|^2$. 
More generally, the full state of an $n$-qubit quantum system is represented with $2^n$ probability amplitudes as
\begin{equation}
  |\psi\rangle = a_{0...00}|0...00\rangle + a_{0...01}|0...01\rangle + ... + a_{1...11}|1...11\rangle
\end{equation}
where $\sum_{i}|a_{i}|^2=1$ \cite{Wu:2019fu}. In quantum computation, the quantum state is updated by applying quantum gates represented in matrix form. For instance, a 1-qubit gate is represented as a 2x2 unitary matrix as follows.
\begin{equation}
  U = \begin{pmatrix}U_{00} & U_{01} \\ U_{10} & U_{11} \\\end{pmatrix}
\end{equation}



State vector simulators store the full quantum state as a vector of $2^n$ probability amplitudes in memory. Thus, the memory space required to store the entire state vector is $2^{n+4}$ bytes when the probability amplitude is double precision. In general, applying a 1-qubit gate $U$ on the $q$-th qubit of an $n$-qubit state vector is represented as repetitive multiplications of the unitary matrix and two-element vectors of probability amplitudes whose indices differ in the $q$-th bits of their binary index \cite{intelqs:2020}:
\begin{equation}
  \begin{pmatrix}
    a'_{*...*0_{q}*...*}\\a'_{*...*1_{q}*...*}
  \end{pmatrix} = 
  \begin{pmatrix}
    U_{00} & U_{01} \\ U_{10} & U_{11} \\
  \end{pmatrix}
  \begin{pmatrix}
    a_{*...*0_{q}*...*}\\a_{*...*1_{q}*...*}
  \end{pmatrix}
  \label{equ:apply_gate}
\end{equation}

A quantum circuit is a computational routine consisting of an ordered sequence of quantum gates, measurements and resets. State vector simulators require an $n$-qubit state vector to simulate an $n$-qubit quantum circuit.

\subsection{Existing Distributed State Vector Simulators}
\label{subsec:simulators}

 Distributed state vector simulators have been developed to simulate large-scale quantum circuits on multiple computing nodes. They can utilize the computational power and memory capacity of multiple nodes by distributing a state vector across them. In this section, we introduce four state-of-the-art distributed simulators: Intel-QS, QuEST, JUQCS-G, and Qiskit Aer.

{\bf Intel-QS} is a CPU-based open-source distributed state vector simulator. It targets public cloud infrastructures as well as HPC systems and is designed for simulation of not only a single large-scale circuit but also multiple small/medium-scale circuits \cite{intelqs:2020}. It is multi-threaded with OpenMP for multi-core CPUs and applies a simple MPI implementation for distributed simulation on multiple nodes. The weak and strong scaling of Intel-QS has been evaluated with up to 42 qubits on 2048 nodes of the {\it SuperMUC-NG} HPC system. Moreover, it is used to investigate the performance of an optimization problem for the quantum approximate optimization algorithm (QAOA).

{\bf QuEST} is an open-source simulator supporting both state vector and density matrix to represent quantum states \cite{quest:2019}. It is multi-threaded with OpenMP and vectorized with SIMD instructions for multi-core CPUs. In addition, it supports MPI-based distributed simulation and GPU acceleration with a single  GPU (but not multiple GPUs). The weak and strong scaling of QuEST has been evaluated using a random circuit consisting of a random sequence of gates with up to 38 qubits on the {\it ARCUS} and {\it ARCHER} supercomputers.

{\bf JUQCS-G} is a GPU-accelerated version of {\it JUQCS} \cite{juqcsg:2021}. JUQCS is a distributed state vector simulator that was used for Google's quantum supremacy demonstration. JUQCS-G supports distributed simulation with multiple GPUs on multiple nodes. It is parallelized with MPI and optimized to reduce the amount of MPI communication by relabeling global and local qubits after a global qubit operation. The weak and strong scaling of JUQCS-G has been evaluated using an Hadamard benchmark circuit (see \refsec{subsec:comparison_juqcsg} for the detail) with up to 40 qubits on 2048 GPUs of the {\it JUWELS Booster} supercomputer. In addition, JUQCS-G is used to study the relationship between quantum annealing and QAOA. Note that it is not an open-source simulator.

{\bf Qiskit Aer} is a set of open-source simulators with realistic noise models for {\it Qiskit}. Qiskit is an open-source framework for working with noisy quantum computers. Qiskit Aer supports multiple types of simulator backends such as state vector and density matrix. In this work, we use the distributed state vector backend of Qiskit Aer that supports MPI-based distributed simulation with multiple GPUs on multiple nodes \cite{qiskitaer:2020}. It applies a cache blocking technique to divide an state vector into small blocks and distribute them across multiple GPUs and multiple nodes. It is also optimized to reduce the amount of MPI communication by appropriately inserting swap gates when a given quantum circuit is transpiled. The weak and strong scaling of this implementation has been evaluated using a {\it Quantum Volume} model circuit (see \refsec{subsec:comparison_qiskitaer} for the detail) with up to 34 qubits on a 16-node IBM cluster system.

\section{mpiQulacs}
\label{sec:mpiqulacs}

We develop a distributed state vector simulator, {\it mpiQulacs}, that is optimized to fully utilize the high memory bandwidth of ARM-based A64FX CPUs and achieve a nearly ideal scalability to larger-scale quantum circuits. As it is based on {\it Qulacs} \cite{qulacs:2021},  we first introduce Qulacs and then explain the three unique features of mpiQulacs in this section.

\subsection{Qulacs}

Qulacs is one of the fastest state vector simulators running on a single machine \cite{qulacs:2021}. It has four main features to accelerate quantum computing research: (1) It is optimized to achieve fast simulation by fully utilizing the capability of multi-core CPUs and GPUs. (2) It is designed to minimize the overhead of pre- and post-processing for fast simulation of small-scale circuits. (3) It is available on several operating systems such as Linux, Windows, and MAC OS with the interfaces for Python and C++ languages. (4) It meets popular demands in quantum computing research by supporting general quantum operations.

However, the scale of quantum circuits Qulacs can simulate is limited by the capability of a single machine, because the original version of Qulacs does not support distributed simulation on multiple computing nodes. In this work, we target A64FX-based cluster systems and extend Qulacs for distributed simulation on them.

\subsection{MPI parallelization}
\label{subsec:mpi_parallelization}

We apply a similar MPI approach as Intel-QS \cite{intelqs:2020} to extend Qulacs for distributed simulation. With $2^p$ MPI processes, the $2^n$ probability amplitudes in an $n$-qubit state vector are evenly distributed to all the processes. In other words, each process stores $2^m\ (m = n - p)$ amplitudes. All quantum operations on the first $m$ qubits require no MPI communication, while it is required when performing operations on the last $p\ (= n - m)$ qubits. Therefore, the qubits with index $0 \leq q < m$ and ones with index $m \leq q < n$ are called {\it local qubits} and {\it global qubits}, respectively.

\begin{figure}[t]
    \centering
    \includegraphics[width=0.49\textwidth]{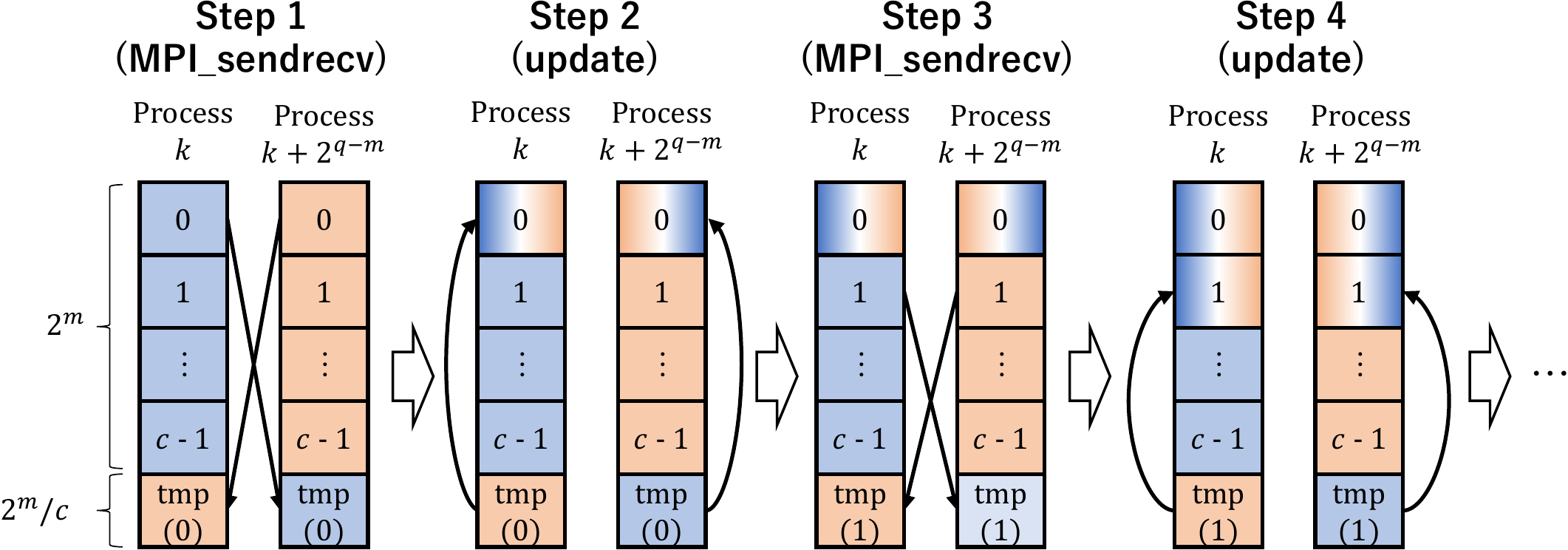}
    \caption{An operation scheme of mpiQulacs on a global qubit. This scheme is performed by all pairs of corresponding processes.}
    \label{fig:mpi_parallelization}
  \end{figure}

  \reffig{fig:mpi_parallelization} illustrates an operation scheme of mpiQulacs on a global qubit. Each process divides the $2^m$ probability amplitudes into fixed-size $c$ chunks and reserves a local temporal buffer to temporarily store one chunk to be exchanged. At the step 1, a pair of a process $k$ and process $k+2^{q-m}$ exchange the first chunks with each other and store them in their temporal buffers. At the step 2, each process updates the first local chunk with the chunk stored in the temporal buffer. These steps are repeated until all of the $c$ chunks are updated. As this operation scheme is performed by all pairs of corresponding processes, $2^{n+4}$-byte MPI communication is performed in total if probability amplitudes are double precision.

\subsection{Vectorization for A64FX CPU}
\label{subsec:tuning_a64fx}

Since we target A64FX CPUs to run mpiQulacs, one of keys for achieving fast simulation is to fully exploit the potential of each A64FX CPU. It is an ARM-based CPU having a 512-bit {\it Scalable Vector Extension~(SVE)} engine~\cite{arm-sve} and 32~GB {\it high-bandwidth memory~(HBM2)}. Thus, we optimize the gate operations of mpiQulacs using this engine to fully utilize the high memory bandwidth of the A64FX CPU.

The gate operations of mpiQulacs are represented as repetitive matrix-vector multiplications as shown in \refequ{equ:apply_gate}. We therefore extend them using 512-bit SVE instructions through the ARM C Language Extension~(ACLE)~\cite{arm2020language}. Since a state vector is represented as an array of double-precision floating point values (i.e., 64-bit values), we optimize each gate operation so that eight values are loaded, processed, and stored simultaneously by an SVE instruction. Consequently, the number of instructions executed in each gate operation is reduced by a factor of eight, and the effective memory bandwidth of each A64FX CPU exceeds 80\% of the peak theoretical bandwidth of HBM2 (see \reffig{fig:QSB_30qubits_1CPU} for the detail).




\subsection{Fused-Swap Gate}
\label{subsec:fused_swap}

\begin{figure}[t]
  \centering
  \includegraphics[width=0.49\textwidth]{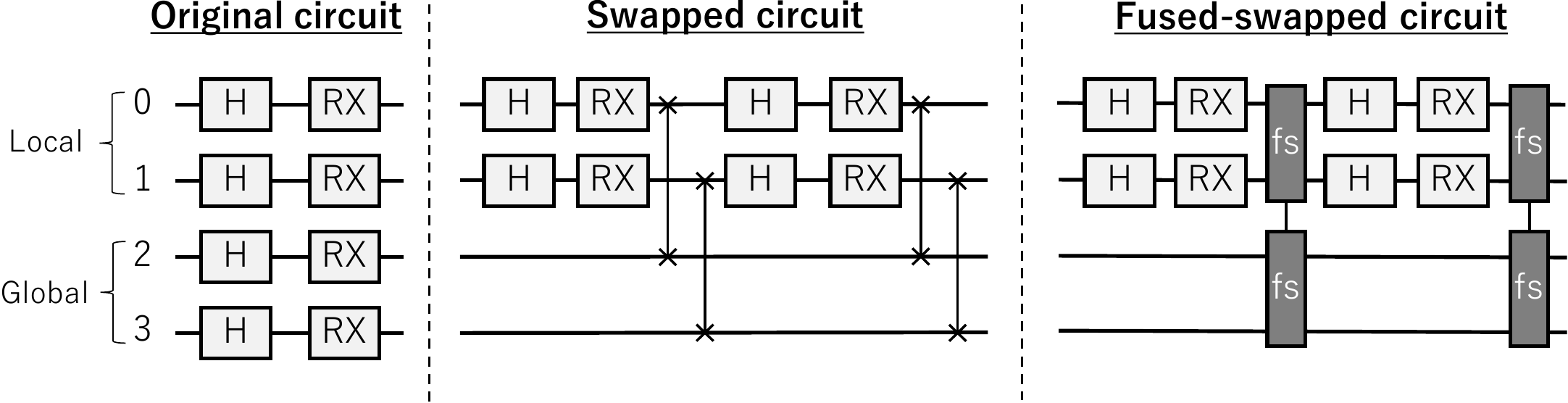}
  \caption{Examples of inserting four swap gates and two fused-swap gates in a 4-qubit circuit with two global qubits.}
  \label{fig:fused_swap}
\end{figure}

A key for achieving a high scalability to larger-scale distributed simulation on tens or hundreds of nodes is to minimize the amount of MPI communication required for global qubit operations as shown in \reffig{fig:mpi_parallelization}, because the inter-node network bandwidth is a bottleneck. One of the straightforward approaches for this goal is to apply gates on as many local qubits as possible by inserting swap gates to a given circuit. A swap gate represented as $swap(i, j)$ is a gate to exchange the probability amplitudes between the $i$-th qubit and $j$-th qubit. If both the qubits are local qubits, the swap operation is performed locally in one process. Otherwise (i.e., if either of them is a global qubit), MPI communication is required. 

\reffig{fig:fused_swap} shows examples of inserting swap gates to a 4-qubit circuit. We here assume that the original circuit applies Hadamard and RX gates on each qubit and the two higher qubits are global qubits, as shown in the left part of this figure. Since a global qubit operation in an $n$-qubit circuit requires $2^{n+4}$-byte MPI communication, the total communication amount for the original circuit including four global qubit operations is 1,024 ($=2^{4+4} \times 4$) bytes. Alternatively, we can apply the eight gates on the two local qubits by inserting four swap gates, as shown in the middle part of \reffig{fig:fused_swap}. As each swap gate involving a global qubit requires $\frac{2^{n+4}}{2}$-byte MPI communication, the four swap gates cause 512-byte ($=\frac{2^{4+4}}{2} \times 4$) MPI communication in total. In this case, the swapped circuit halves the total amount of MPI communication compared to the original circuit. The distributed implementation of Qiskit Aer introduced in \refsec{subsec:simulators} applies this approach \cite{qiskitaer:2020}. However, an important note is that the total amount of MPI communication is ($\frac{2^{n+4}}{2} \times \#swap\_gates$) bytes in general, which increases linearly as the number of swap gates is increased.

In order to further reduce the amount of MPI communication, we implement a {\it fused-swap} gate in mpiQulacs that collectively performs multiple swap operations as one swap operation. It is based on the idea discussed in \cite{Raedt:2006ma} and designed to fully utilize the bandwidth of InfiniBand.  Provided that $s$ is the number of swap operations to be fused, a fused-swap operation behaves in a similar way to the following code:
\begin{lstlisting}[label=prm1, caption=Fused-swap behavior]
fused_swap(p, q, s){
    for(i=0; i<s; i++){
        swap(p+i, q+i);
    }
}
\end{lstlisting}
As shown in this code, the fused-swap gate contiguously performs $s$ swap operations from the $p$-th and $q$-th qubits. As it does not allow the two operation ranges ($[p:p+s]$ and $[q:q+s]$) to overlap, each swap operation can be performed independently. 

A fused-swap operation exchanges the probability amplitudes within the two operation ranges in a state vector between two corresponding processes at three steps: (1) A {\it gather operation} gathers a block of amplitudes into the send buffer of a source process. (2) The block stored in the send buffer is sent to the receive buffer of a target process with MPI, while the receive buffer of the source process receives a block sent from the send buffer of the target process. (3) A {\it scatter operation} scatters the amplitudes stored in the receive buffer to appropriate positions in a state vector. These three steps are repeated until all of the amplitudes within the two operation ranges are exchanged with all corresponding target processes.

The total amount of MPI communication in a circuit including fused-swap gates is ($2^{n+4} \times (1 - \frac{1}{2^s}) \times \#fused\_swap\_gates$) bytes. The right part of \reffig{fig:fused_swap} shows an example of inserting two fused-swap gates to the 4-qubit circuit, where the total amount of MPI communication is 384 bytes ($=2^{4+4} \times (1 - \frac{1}{2^2}) \times 2$). It is 25\% lower than that of the swapped circuit. Although the examples in \reffig{fig:fused_swap} are tiny, the benefit of the fused-swap gate is more significant for larger-scale circuits. If $s$ is large (e.g., 4 or more) and a fused-swap gate fuses $s$ swap gates, we can bound the total amount of MPI communication to $\approx 2^{n+4} \times \frac{\#swap\_gates}{s}$ bytes.

\subsection{Double Buffering for Fused-Swap}
\label{subsec:double_buffering}

\begin{figure}[t]
    \centering
    \includegraphics[width=0.9\columnwidth]{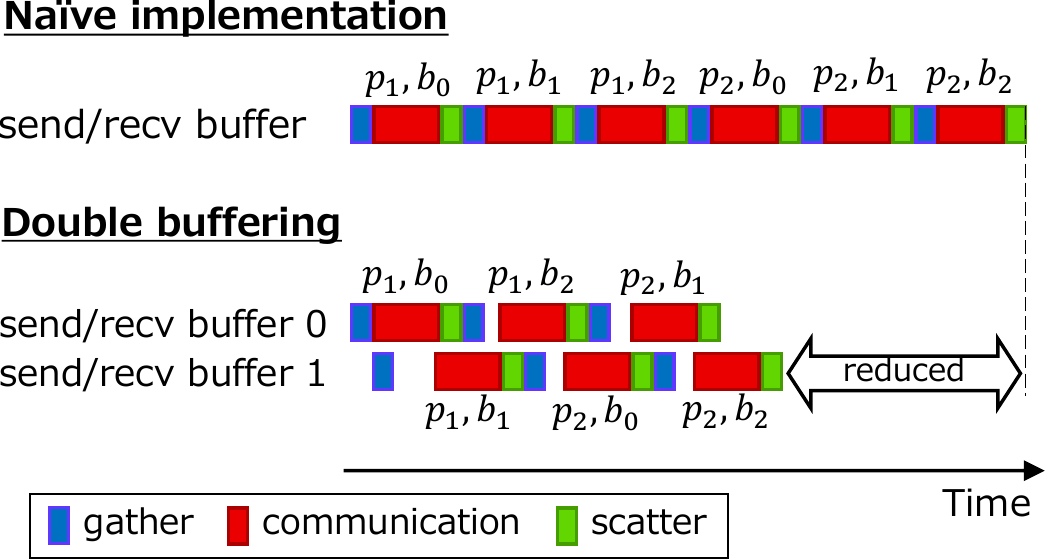}
    \caption{An example of double buffering for a fused-swap operation. A pair of $p_i$ and $b_j$ represents a $j$-th buffer procedure for an $i$-th target process.}
    \label{fig:double_buffering}
  \end{figure}

A na\"ive implementation of a fused-swap operation sequentially performs the procedure of a gather operation, MPI communication, and a scatter operation, as shown in the upper part of \reffig{fig:double_buffering}. To reduce the processing time of a fused-swap operation, we apply {\it double buffering}, which is a well-known optimization technique to overlap computation and communication. The lower part of \reffig{fig:double_buffering} illustrates our double buffering implementation. We prepare two pairs of send/receive buffers and overlap the $(j-1)$-th scatter operation and the $(j+1)$-th gather operation with $j$-th MPI communication. In addition, we also overlap the last scatter operation of the $i$-th target process and the first MPI communication of the $(i+1)$-th target process. Consequently, the double buffering implementation reduces the processing time of a fused-swap operation by hiding the processing times of gather and scatter operations.

\section{Evaluation}
\label{sec:evaluation}

In this section, we evaluate the performance of mpiQulacs on our 64-node A64FX-based cluster system named {\it Todoroki} and compare the results with those of Intel-QS, JUQCS-G, and Qiskit Aer reported in corresponding papers. Moreover, we evaluate the performance of three open-source distributed simulators (Intel-QS, QuEST, and Qiskit Aer) by ourself on the {\it ABCI} supercomputer and compare those results with that of mpiQulacs evaluated on Todoroki. We first explain the configuration of Todoroki and evaluation metrics and then show the comparison results.

\subsection{Todoroki: 64-node A64FX-based cluster system}
\label{subsec:todoroki}

We construct a new 64-node A64FX-based cluster system named {\it Todoroki} to evaluate mpiQulacs. \reftab{tab:fx700} summarizes its configuration. It consists of 64 {\it PRIMEHPC FX700} servers, each of which has a 48-core ARM-based A64FX CPU. This CPU is also used in the world's top Fugaku supercomputer. The noticeable feature of this CPU is 32~GB high-bandwidth memory (HBM2), whose theoretical peak bandwidth is 1,024 GB/s. This is an order of magnitude higher than that of widely used DDR4 DRAM. All of the 64 nodes are interconnected with InfiniBand EDR in a fat tree topology.

\begin{table}[ht]
  \caption{Configuration of Todoroki}
  \begin{center}
  \begin{tabular}{l|l}
      \hline
      Cluster node & PRIMEHPC FX700 \\
      \# of nodes & 64 \\
      \hline \hline
      CPU & ARM-based A64FX \\
      \# of CPUs per node & 1 \\
      \# of cores per CPU & 48 \\
      CPU memory & 32~GB HBM2 \\
      Theoretical peak FLOPS per CPU & 3.1~TFLOPS \\
      Theoretical peak memory BW per CPU & 1,024~GB/s \\
      \hline
      Interconnect & InfiniBand EDR \\
      \# of HCAs per node & 1 \\
      Theoretical peak network BW per node & 12.5~GB/s \\
      \hline
      OS & CentOS 8 \\
      Compiler & GCC 11.2.1 \\
      MPI & OpenMPI 4.1.2 \\
      \hline
  \end{tabular}
  \label{tab:fx700}
  \end{center}
\end{table}

\subsection{Evaluation Metrics}

To compare mpiQulacs with other existing distributed simulators, we measure the execution times of several quantum benchmark circuits. Note that each circuit is executed six times and we report the average execution time of the last five runs (i.e., the first run is omitted). The detail of each circuit is described in each of the following  sections.

In addition, we define a novel metric, {\it quantum Q/F ratio (QBF)}, that indicates the execution efficiency of state vector simulation running on cluster systems and use it to compare mpiQulacs running on Todoroki and the existing simulators running on HPC systems. It is defined as
\begin{equation}
  Quantum\ B/F\ ratio = \frac{2^{n+5} \times \#gates}{exetime \times totalFLOPS}
\end{equation}
where $n$, $\#gates$, and $exetime$ are the number of qubits, the number of gates, and the execution time of a given quantum circuit, respectively. $totalFLOPS$ is the total theoretical peak FLOPS of all computing units (i.e., CPUs or GPUs) used to execute the circuit. According to \cite{qhipster:2016}, $2^{n+5}$ represents the amount of memory traffic in bytes to read and write an entire state vector in one gate operation. Thus, the product of $2^{n+5}$ and $\#gates$ corresponds to the total amount of memory traffic in bytes for the circuit. On the other hand, the product of $exetime$ and $totalFLOPS$ corresponds to the estimated number of double-precision floating point operations executed by CPUs or GPUs for the target circuit. The QBF is inspired by {\it B/F ratio}, which is a metric widely used for HPC systems, but has a different meaning. A higher QBF means that a larger-scale quantum circuit can be simulated in a shorter time with less computational hardware resource.

\subsection{Comparison with Intel-QS}

We measure the processing time of a 1-qubit gate of mpiQulacs and compare it with that of Intel-QS reported in \cite{intelqs:2020}. The weak and strong scaling of Intel-QS has been evaluated on the 6,480-node SuperMUC-NG HPC system. \reftab{tab:supermuc-ng} summarizes the configuration of this system. Each node has two sockets of Xeon\textregistered\ Platinum 8174 processors and 96~GB DDR4 DRAM. We calculate the theoretical peak FLOPS of each CPU by dividing the total theoretical peak FLOPS of the entire system by the total number of CPUs (= 26.3 PFLOPS / 12,960 CPUs). Moreover, we calculate the theoretical peak memory bandwidth based on the memory type and the number of memory channels supported by the CPU (= 2666 MHz $\times$ 8 bytes $\times$ 6 channels).

\begin{table}[ht]
  \caption{Configuration of SuperMUC-NG}
  \begin{center}
  \begin{tabular}{l|l}
      \hline
      \# of nodes & 6,480 \\
      \hline \hline
      CPU & Xeon\textregistered\ Platinum 8174 \\
      \# of CPUs per node & 2 \\
      \# of cores per CPU & 24 \\
      CPU memory & 96~GB DDR4 DRAM \\
      Theoretical peak FLOPS per CPU & 2.0 TFLOPS \\
      Theoretical peak memory BW per CPU & 125 GB/s \\
      \hline
      Interconnect & 100~Gbit/s OmniPath \\
      Theoretical peak network BW per node & 12.5~GB/s \\
      \hline
      OS & Suse Linux \\
      \hline
  \end{tabular}
  \label{tab:supermuc-ng}
  \end{center}
\end{table}

\begin{figure}[t]
  \centering
  \subfloat[Weak scaling (30 qubits per CPU)]
  {\includegraphics[width=0.48\textwidth]{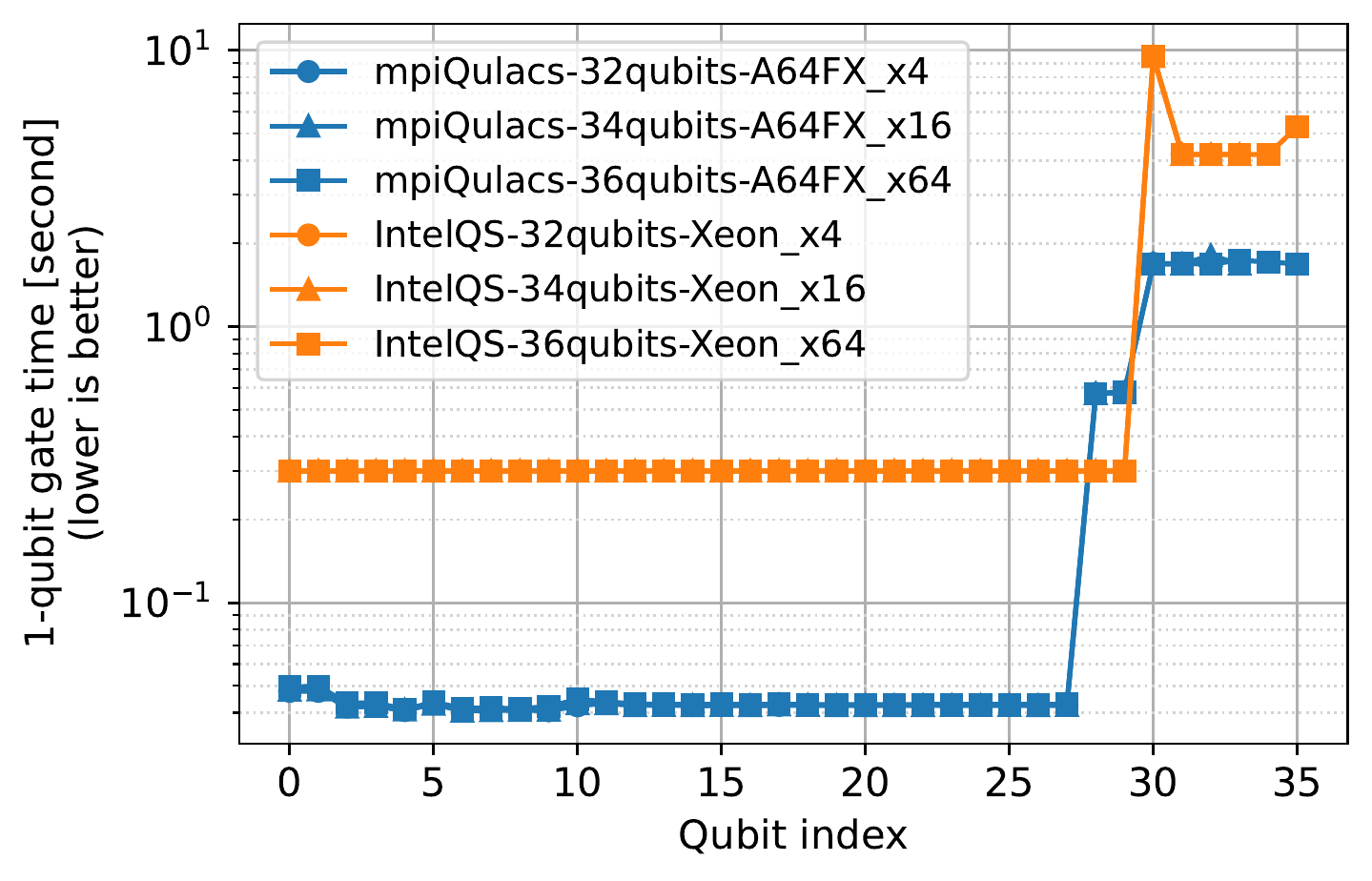}%
  \label{subfig:weak_scaling_1qbgate_time}}
  \\
  \subfloat[Strong scaling (fixed 32 qubits)]
  {\includegraphics[width=0.48\textwidth]{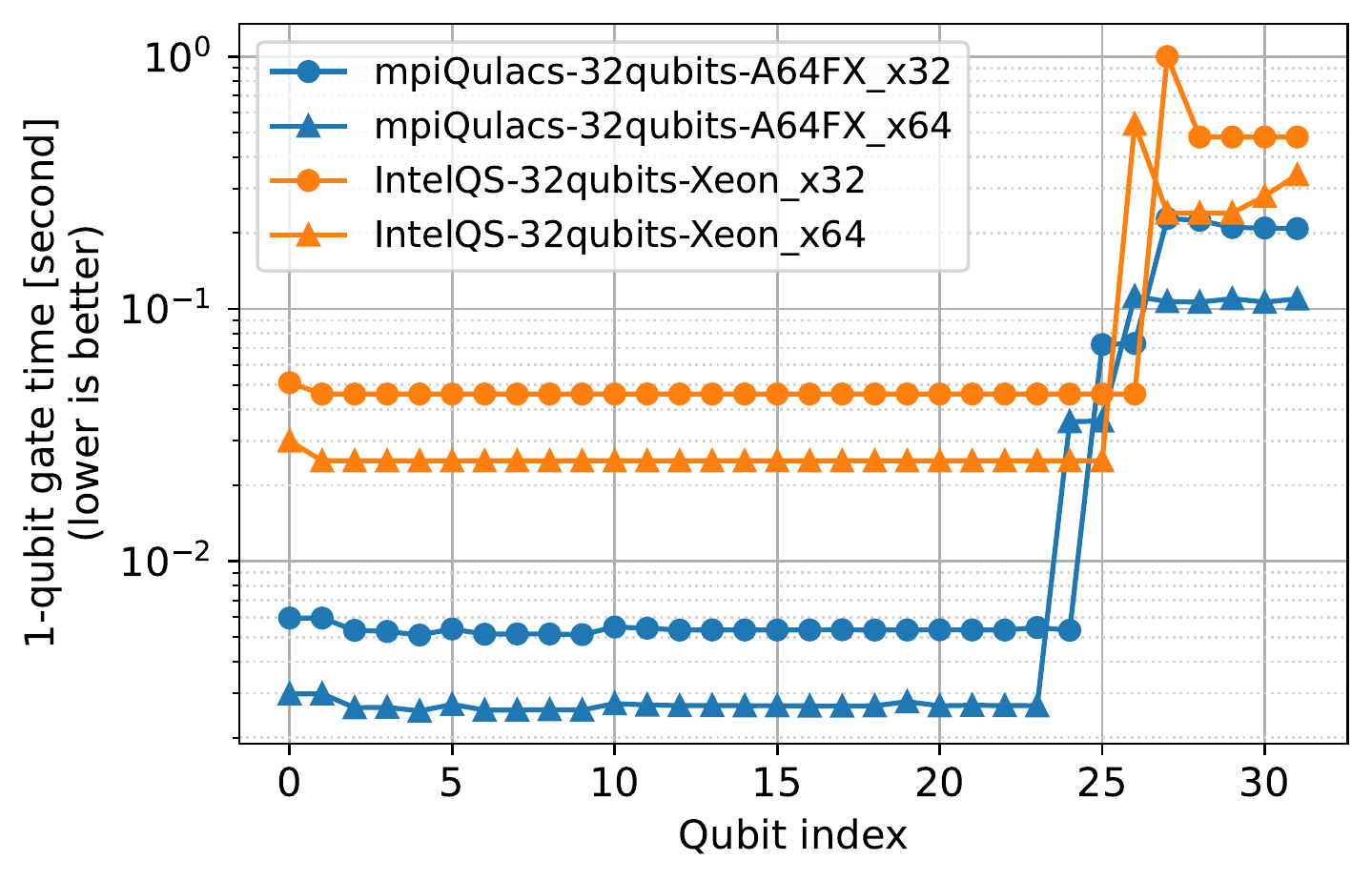}%
  \label{subfig:strong_scaling_1qbgate_time}}
  \caption{The 1-qubit gate time on each qubit of mpiQulacs running on Todoroki and Intel-QS running on SuperMUC-NG.}
  \label{fig:comparison_intelqs}
\end{figure}

The 1-qubit gate time of Intel-QS has been measured using a 1-qubit gate defined by a random 2x2 matrix \cite{intelqs:2020}. We therefore measure the processing time of an RX-gate, which is also defined by a 2x2 matrix, with mpiQulacs running on Todoroki. \reffig{fig:comparison_intelqs} plots the 1-qubit gate time when applied on each qubit with mpiQulacs and Intel-QS.

\reffig{subfig:weak_scaling_1qbgate_time} shows the weak scaling results of both the simulators, where the number of qubits is set to 30 per CPU and incremented with doubling the number of CPUs. Both mpiQulacs and Intel-QS achieve ideal weak scaling because the 1-qubit gate time is not changed (three blue and orange lines completely overlap) even when the number of CPUs is 4x increased. The left floor parts of blue and orange lines show the processing time of a local qubit, which requires no MPI communication. Here, mpiQulacs achieves about 8x higher performance than Intel-QS by fully utilizing the high memory bandwidth of A64FX CPUs. On the other hand, the right stair parts of blue and orange lines show the processing time of a global qubit, which requires MPI communication. The performance of mpiQulacs is here almost twice as high as that of Intel-QS. This performance difference can be derived from the difference of the total network bandwidth that depends on the number of nodes used, because mpiQulacs uses the twice number of nodes compared to Intel-QS.

\reffig{subfig:strong_scaling_1qbgate_time} shows the strong scaling results, where 32 or 64 CPUs are used to process the 1-qubit gate with fixed 32 qubits. Both the simulators also achieve nearly ideal strong scaling because the 1-qubit gate time is almost halved by doubling CPU counts. The performance difference between them is almost equivalent to that shown in \reffig{subfig:weak_scaling_1qbgate_time}.

\begin{figure}[t]
  \centering
  \includegraphics[width=0.48\textwidth]{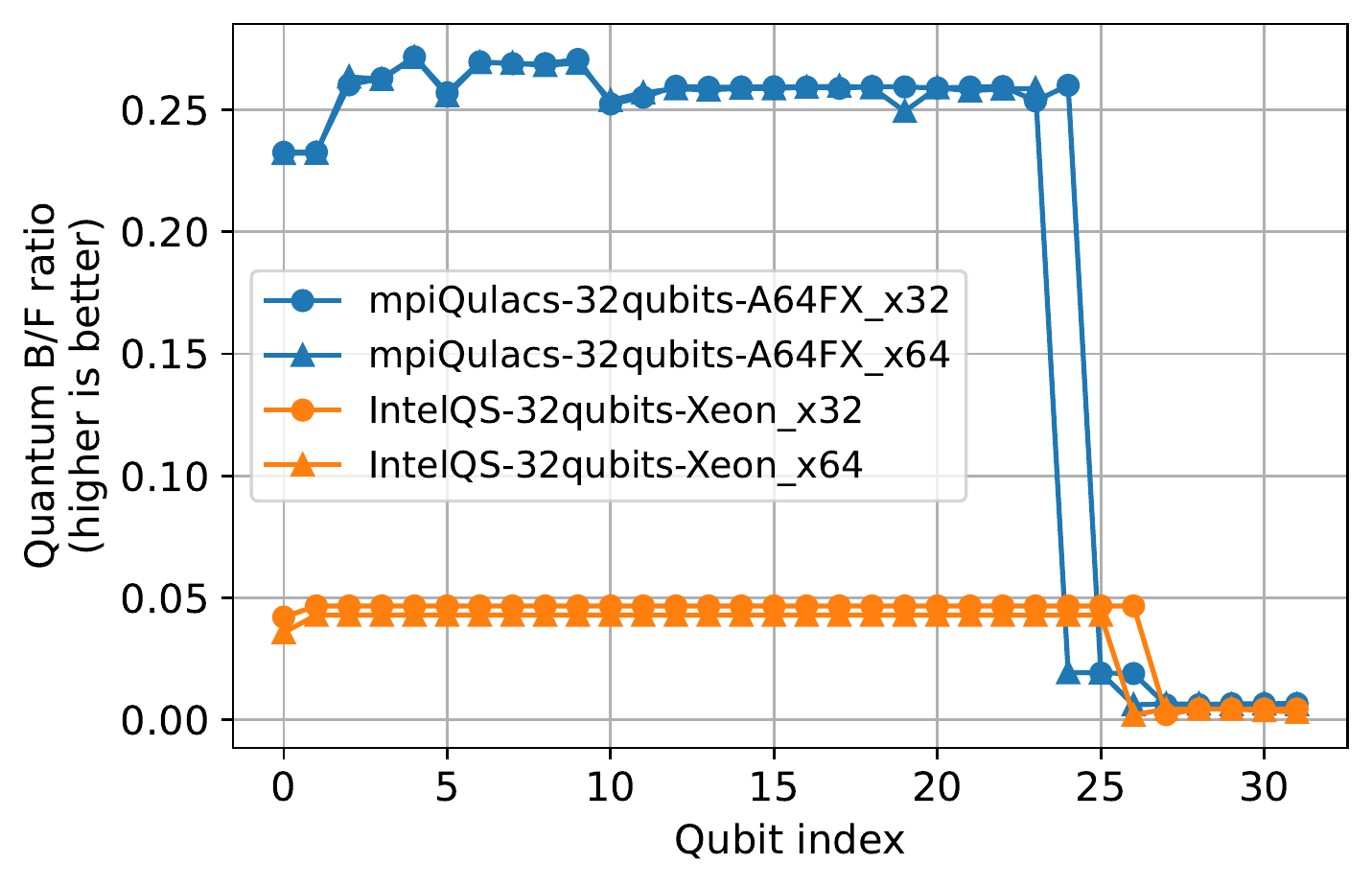}
  \caption{Quantum B/F ratio based on the strong scaling results of mpiQulacs running on Todoroki and Intel-QS running on SuperMUC-NG.}
  \label{fig:qbf_intelqs}
\end{figure}

Moreover, \reffig{fig:qbf_intelqs} plots the QBF of both the simulators based on the strong scaling results  shown in \reffig{subfig:strong_scaling_1qbgate_time}. mpiQulacs achieves over 5x higher QBF than Intel-QS for a local qubit operation, because it achieves 8x higher performance with 1.55x higher total FLOPS. However,  both the simulators drop the QBF significantly for a global qubit operation, because the 1-qubit gate time is significantly increased by MPI communication. The reason why the QBF of both the simulators is not significantly changed even when the number of CPUs is doubled is that the 1-qubit gate time is halved by doubling the total FLOPS. 

{\it Summary of comparison with Intel-QS}: Both mpiQulacs and Intel-QS achieve nearly ideal weak and strong scaling. mpiQulacs significantly outperforms Intel-QS especially for a local qubit operation by fully utilizing the high memory bandwidth of A64FX CPUs. This significant performance improvement also brings the much higher QBF of mpiQulacs.

\subsection{Comparison with JUQCS-G}
\label{subsec:comparison_juqcsg}

We then compare mpiQulacs with JUQCS-G evaluated on the JUWELS Booster supercomputer. \reftab{tab:juwels_booster} summarizes the configuration of the system. It consists of 936 nodes, each of which has two sockets of AMD EPYC\texttrademark\ 7402 processors with 512~GB DDR4 DRAM and four NVIDIA A100 GPUs. This GPU contains 40~GB HBM2 as well as the A64FX CPU, and its theoretical peak memory bandwidth is 1,555 GB/s. The major difference of the GPU from the A64FX CPU is over 6x higher FLOPS.

\begin{table}[ht]
  \caption{Configuration of JUWELS Booster}
  \begin{center}
  \begin{tabular}{l|l}
      \hline
      \# of nodes & 936 \\
      \hline \hline
      CPU & AMD EPYC\texttrademark\ 7402 \\
      \# of CPUs per node & 2 \\
      \# of cores per CPU & 24 \\
      CPU memory & 512~GB DDR4 DRAM \\
      \hline
      GPU & NVIDIA A100 \\
      \# of GPUs per node & 4 \\
      GPU memory & 40~GB HBM2 \\
      Theoretical peak FLOPS per GPU & 19.5 TFLOPS \\
      Theoretical peak memory BW per GPU & 1,555 GB/s \\
      GPU interconnect & NVLink\textregistered\ 3 \\
      \hline
      Interconnect & InfiniBand HDR \\
      \# of HCAs per node & 4 \\
      Theoretical peak network BW per node & 100~GB/s \\
      \hline
      OS & CentOS\\
      \hline
  \end{tabular}
  \label{tab:juwels_booster}
  \end{center}
\end{table}

\begin{figure}[t]
  \centering
  \subfloat[Weak scaling  (30 qubits per CPU and 31 qubits per GPU)]
  {\includegraphics[width=0.48\textwidth]{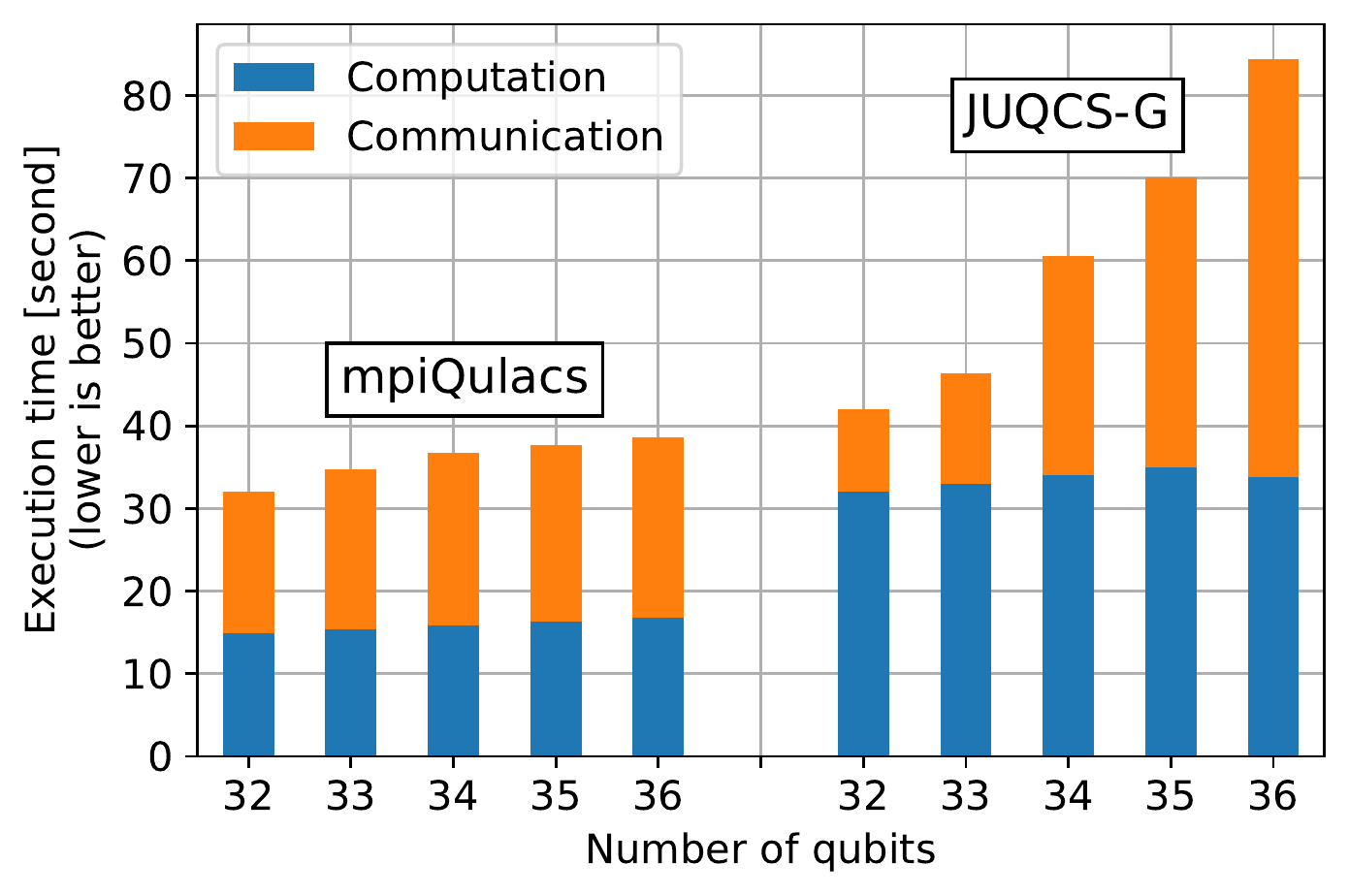}%
  \label{subfig:weak_scaling_Hx11_time}}
  \\
  \subfloat[Strong scaling (fixed 34 qubits)]
  {\includegraphics[width=0.48\textwidth]{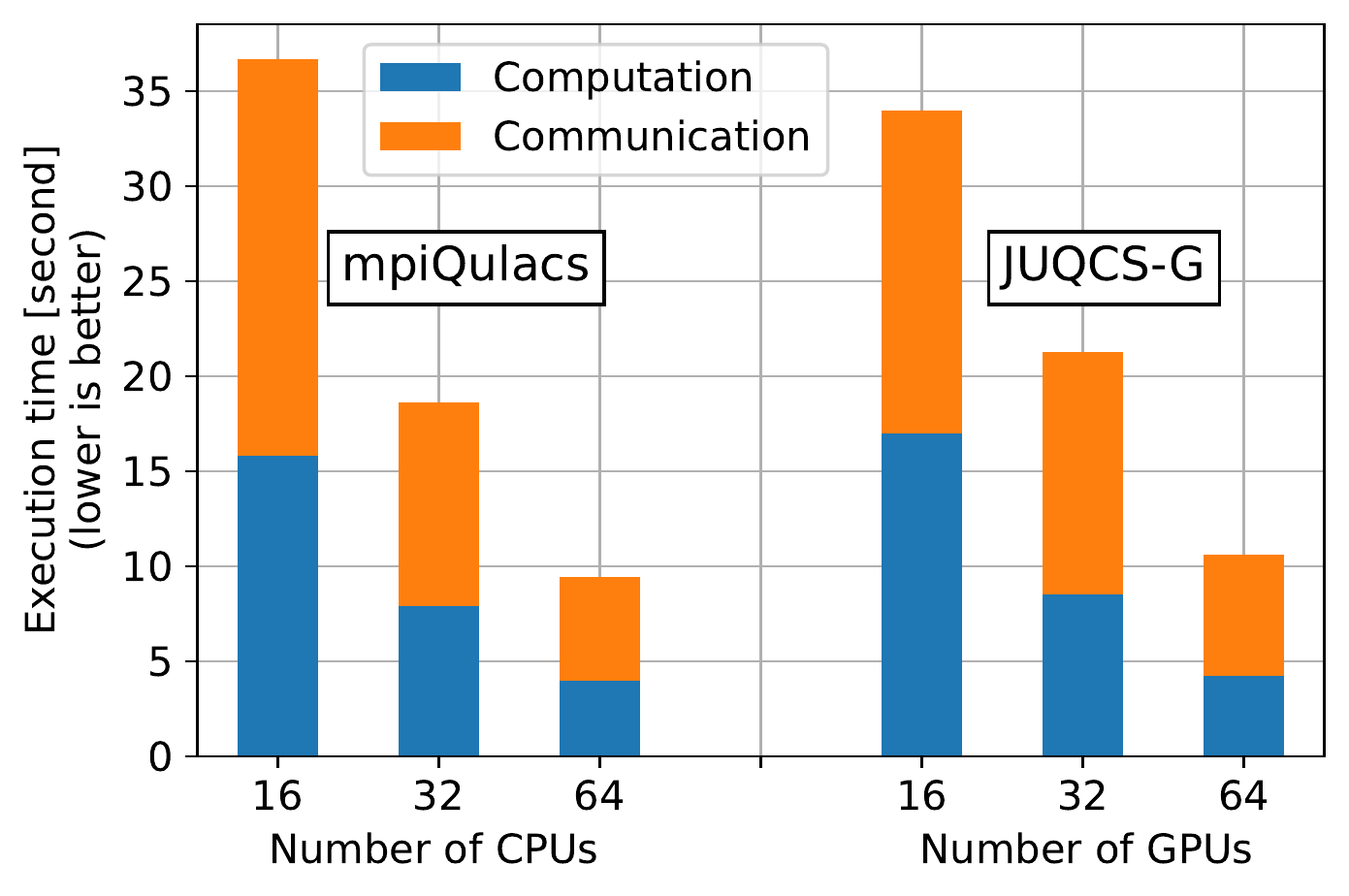}%
  \label{subfig:strong_scaling_Hx11_time}}
  \caption{The execution time of Hadamard benchmark circuit with mpiQulacs running on Todoroki and JUQCS-G running on JUWELS Booster.}
  \label{fig:comparison_juqcsg}
\end{figure}

The weak and strong scaling of JUQCS-G has been evaluated using an Hadamard benchmark circuit \cite{juqcsg:2021}, where eleven Hadamard gates are repeatedly applied on each qubit. Thus, we also implement it for mpiQulacs. \reffig{fig:comparison_juqcsg} plots the computation time and communication time for this circuit with mpiQulacs running on Todoroki and JUQCS-G running on JUWELS Booster. We obtain the results of JUQCS-G from \cite{juqcsg:2021}, where the normalized execution time with respect to the number of gates in the 32-qubit circuit is reported. We therefore calculate the absolute execution time by multiplying the normalized time by a corresponding normalization factor and plot it in \reffig{fig:comparison_juqcsg}. For instance, the normalization factor of the 36-qubit circuit is 1.125 (=$\frac{36\times11}{32\times11}$). For mpiQulacs, we insert fused-swap gates that involves all global qubits (i.e., $s = \#global\_qubits$) in the circuit so that the amount of MPI communication is minimized and the final state vector is consistent to the original circuit. Note that the communication time of mpiQulacs includes the processing time of fused-swap gates in addition to the MPI communication time.

\begin{table}[ht]
  \caption{The configurations of Todoroki and JUWELS Booster for the 36-qubit Hadamard benchmark circuit}
  \begin{center}
  \begin{tabular}{l|r|r}
      \hline
      & \multicolumn{1}{|c|}{Todoroki}  & \multicolumn{1}{|c}{JUWELS Booster} \\
      \hline \hline
      \# of nodes & 64 & 8 \\
      \# of CPUs or GPUs & 64 CPUs & 32 GPUs \\
      Total theoretical FLOPS & 198 TFLOPS & 624 TFLOPS (x3.15)\\
      Total theoretical memory BW & 64 TB/s & 49 TB/s (x0.76)\\
      Total theoretical network BW & 800 GB/s & 800 GB/s (x1.00)\\
      \hline
      \multicolumn{3}{l}{* The values in () indicates the related values to Todoroki.} \\
  \end{tabular}
  \end{center}
  \label{tab:todoroki_juwels_36qubits}
\end{table}

\reffig{subfig:weak_scaling_Hx11_time} shows the weak scaling results of both the simulators. The maximum numbers of qubits on one A64FX CPU and one A100 GPU are 30 and 31, respectively. We therefore start at 32 qubits with four A64FX CPUs for mpiQulacs and with two A100 GPUs for JUQCS-G and increment the number of qubits up to 36 while doubling the number of CPUs or GPUs. \reftab{tab:todoroki_juwels_36qubits} summarizes the configurations of Todoroki and JUWELS Booster when the number of qubits is 36 and shows that the theoretical performance of Todoroki is not significantly different from that of JUWELS Booster except the x3.15 lower total FLOPS.  We can see in \reffig{subfig:weak_scaling_Hx11_time} that mpiQulacs outperforms JUQCS-G consistently across the varied number of qubits. Interestingly, the performance difference between them gets more significant as the number of qubits is increased and reaches to over twice with 36 qubits. The fused-swap gates of mpiQulacs mainly contribute to this result by minimizing the amount of MPI communication. In fact, the communication time of mpiQulacs does not increase linearly with the increased number of qubits. On the other hand, since JUQCS-G requires MPI communication for all global qubit operations, its communication time increases linearly. The computation time of mpiQulacs is almost half of that of JUQCS-G even though the total FLOPS for mpiQulacs is 3.15x lower than that for JUQCS-G. This result implies that state vector simulation is memory bound, and the high memory bandwidth of the A64FX fits this requirement.

\reffig{subfig:strong_scaling_Hx11_time} shows the strong scaling results, where the number of CPUs or GPUs is increased from 16 to 64 with fixed 34 qubits. The computation time of mpiQulacs is comparable to that of JUQCS-G when the numbers of CPUs and GPUs are same. This result again demonstrates that high memory bandwidth is a key for fast state vector simulation.

\begin{figure}[t]
  \centering
  \includegraphics[width=0.48\textwidth]{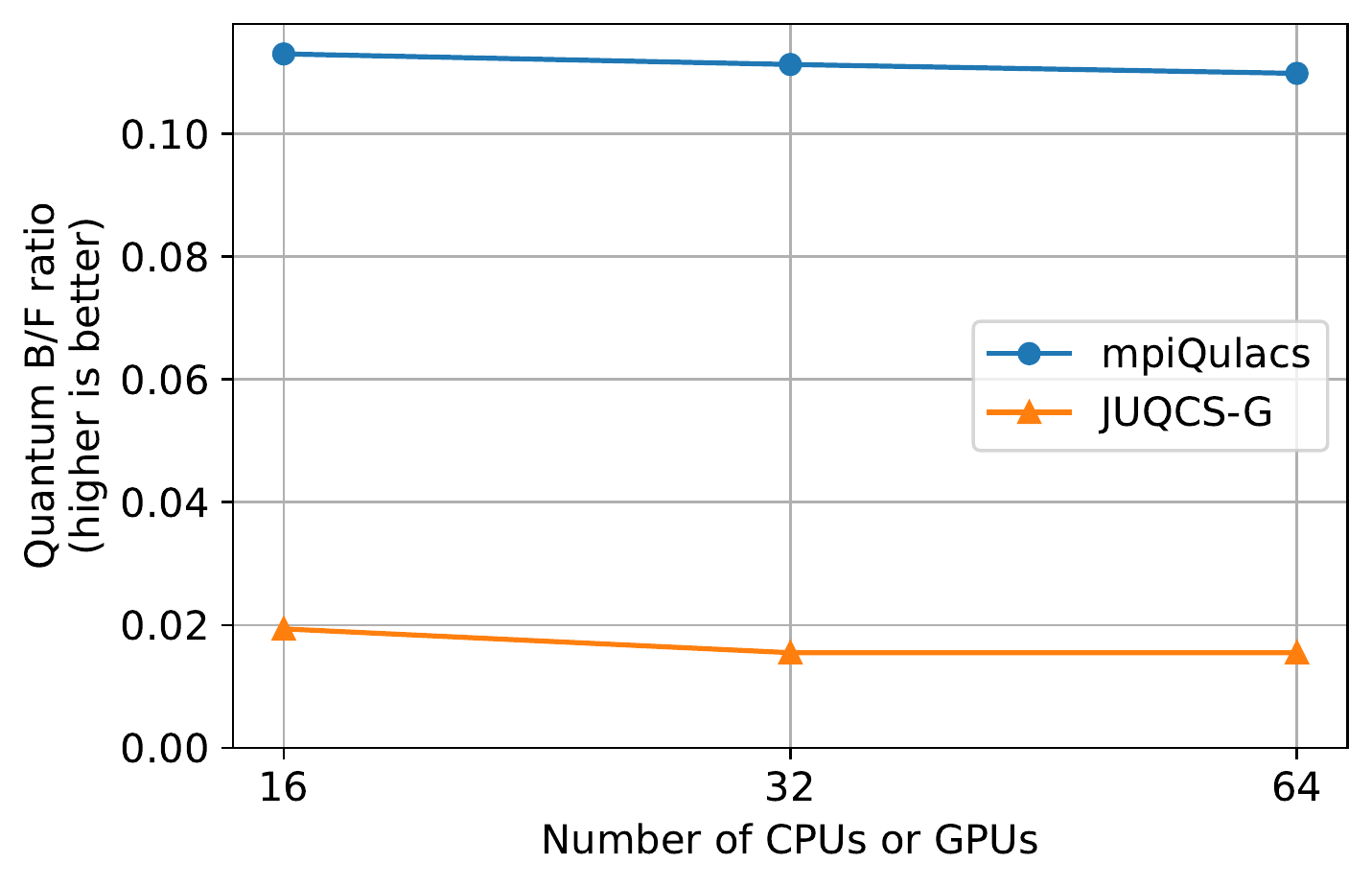}
  \caption{Quantum B/F ratio based on the strong scaling results of mpiQulacs running on Todoroki and JUQCS-G running on JUWELS Booster.}
  \label{fig:strong_scaling_Hx11_qbf}
\end{figure}

\reffig{fig:strong_scaling_Hx11_qbf} plots the QBF of both the simulators based on the strong scaling results shown in \reffig{subfig:strong_scaling_Hx11_time}. mpiQulacs achieves an order of magnitude higher QBF than JUQCS-G. This is because the FLOPS of one A64FX CPU is over 6x lower than that of one A100 GPU, whereas the execution time is comparable. This result means that the FLOPS of the A64FX CPU is sufficient for state vector simulation. The QBF of both the simulators is almost constant because the execution time is halved by doubling the number of CPUs or GPUs.

{\it Summary of comparison with JUQCS-G}: mpiQulacs outperforms JUQCS-G especially for large-scale simulation by minimizing MPI communication. In addition, mpiQulacs enables one A64FX CPU to achieve comparable performance to one A100 GPU, leading to an order of magnitude higher QBF. 

\subsection{Comparison with Qiskit Aer}
\label{subsec:comparison_qiskitaer}

Next, we compare mpiQulacs with the distributed state vector backend of Qiskit Aer that supports distributed simulation with multiple GPUs on multiple nodes \cite{qiskitaer:2020} as introduced in \refsec{subsec:simulators}. The weak and strong scaling of this backend has been evaluated on a 16-node IBM GPU cluster system. \reftab{tab:ibm_cluster} summarizes its configuration. Each node has two sockets of Power9 processors with 512~GB DDR4 DRAM and six NVIDIA V100 GPUs. This GPU contains 16~GB HBM2, whose theoretical peak bandwidth is 900~GB/s.

\begin{table}[ht]
  \caption{Configuration of IBM GPU cluster system}
  \begin{center}
  \begin{tabular}{l|l}
      \hline
      Cluster node & IBM Power System AC922 \\
      \# of nodes & 16 \\
      \hline \hline
      CPU & POWER9 \\
      \# of CPUs per node & 2 \\
      \# of cores per CPU & 21 \\
      CPU memory & 512~GB DDR4 DRAM\\
      \hline
      GPU & NVIDIA V100 \\
      \# of GPUs per node & 6 \\
      GPU memory & 16~GB HBM2 \\
      Theoretical peak FLOPS per GPU & 7.0 TFLOPS \\
      Theoretical peak memory BW per GPU & 900 GB/s \\
      GPU interconnect & NVLink\textregistered\ 2 \\
      \hline
      Interconnect & InfiniBand EDR \\
      Theoretical peak network BW per node & 12.5~GB/s \\
      \hline
      OS & RHEL Server 7.6 \\
      Compiler & GCC 8.3.0 \\
      CUDA Toolkit & CUDA 10.1 \\
      MPI & IBM Spectrum MPI 10.3.1 \\
      \hline
  \end{tabular}
  \label{tab:ibm_cluster}
  \end{center}
\end{table}

\begin{figure}[t]
  \centering
  \subfloat[Weak scaling (30 qubits per node)]
  {\includegraphics[width=0.48\textwidth]{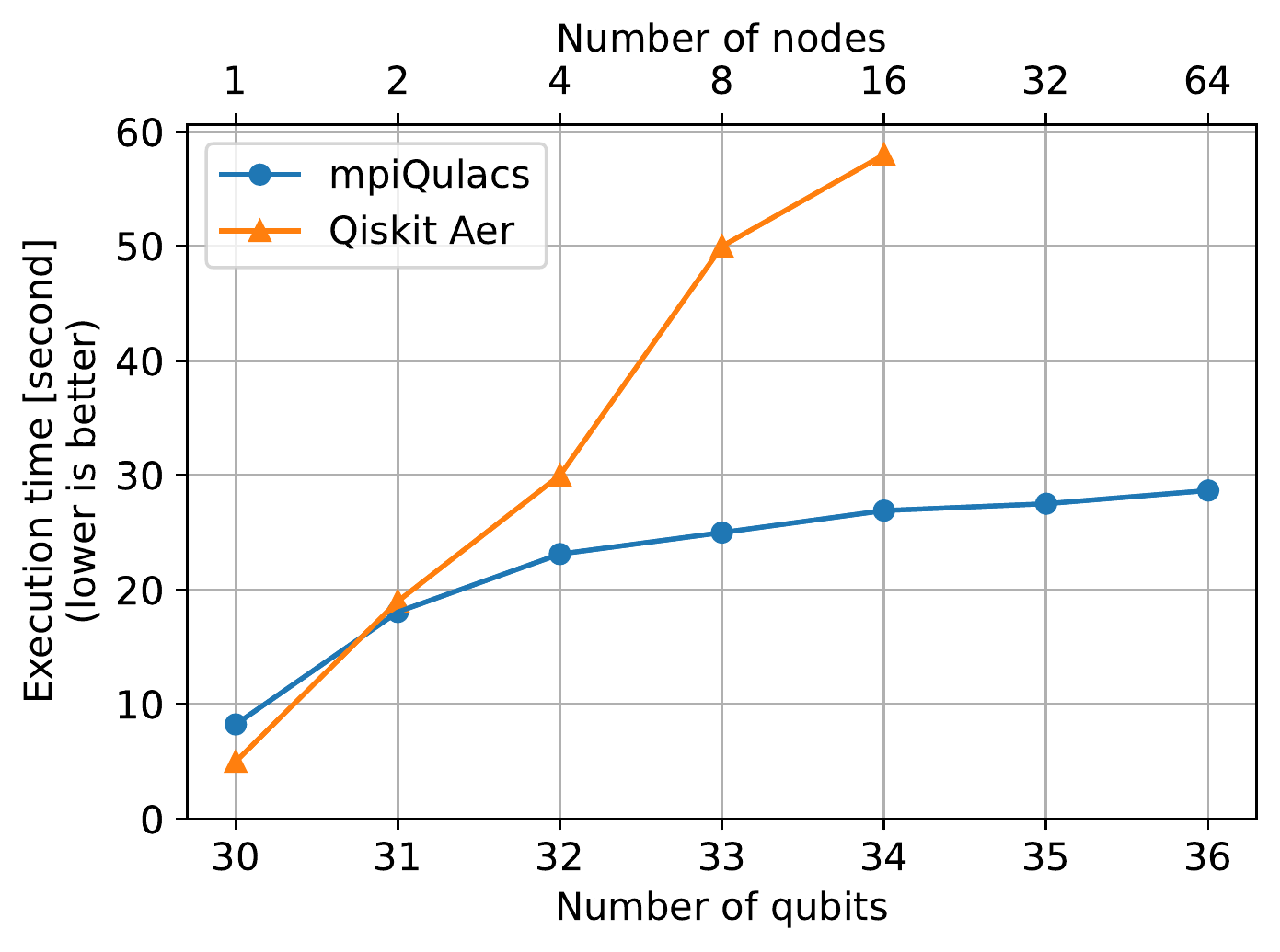}%
  \label{subfig:weak_scaling_QV_time}}
  \\
  \subfloat[Strong scaling (fixed 30 qubits)]
  {\includegraphics[width=0.48\textwidth]{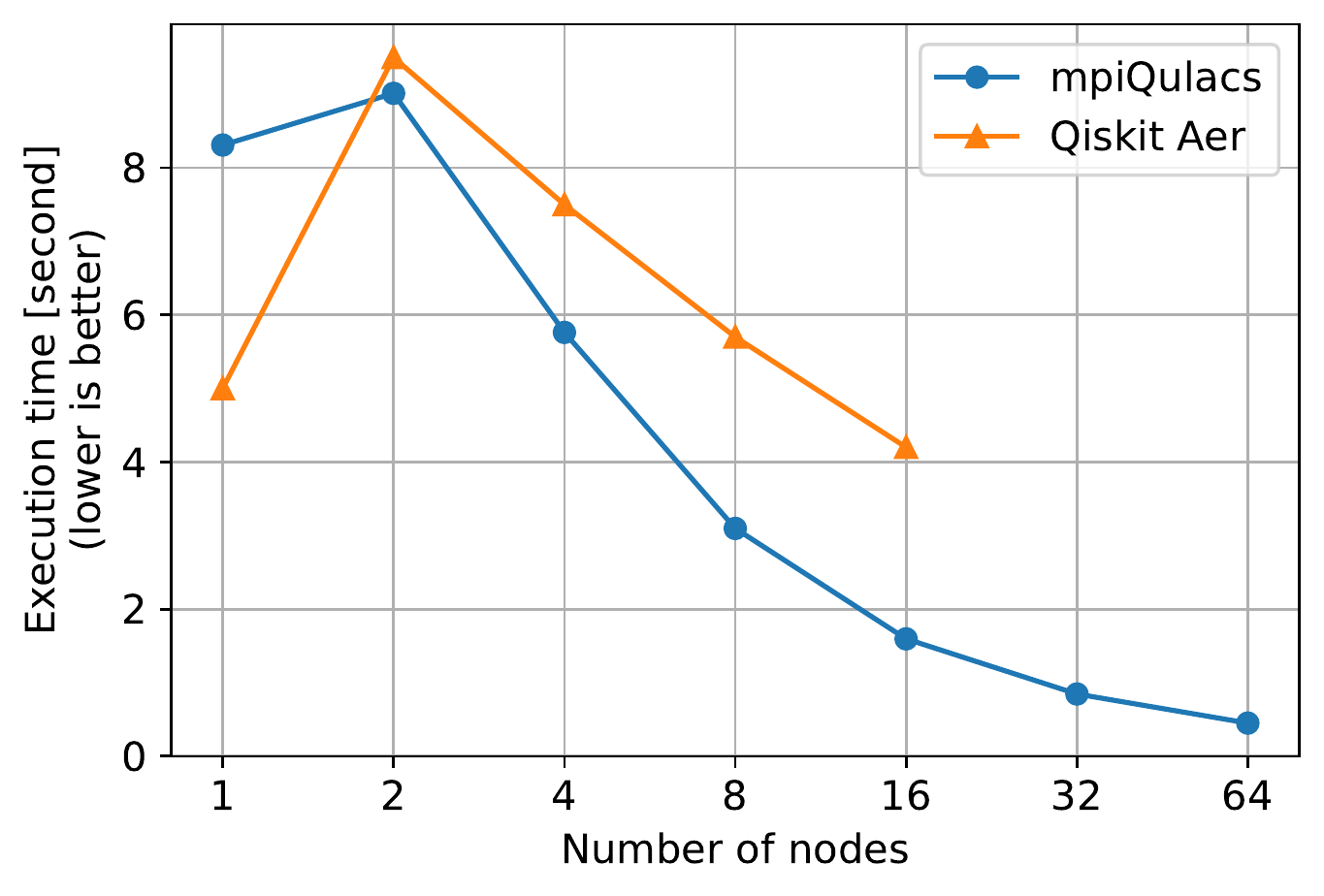}%
  \label{subfig:strong_scaling_QV_time}}
  \caption{The execution time of Quantum Volume model circuit with mpiQulacs running on Todoroki and Qiskit Aer running on an IBM GPU cluster system.}
  \label{fig:comparison_qiskitaer}
\end{figure}

A {\it Quantum Volume} model circuit has been used for the evaluation of Qiskit Aer \cite{qiskitaer:2020}. Quantum Volume (QV) is a single-number metric to measure the performance of near-term quantum computers \cite{quantum_volume:2019}. The model circuit for QV measurement consists of multiple layers of random permutations of qubit labels, followed by random 2-qubit dense-matrix gates. We implement this circuit for mpiQulacs according to the source code on Github \cite{qv_source} and insert fused-swap gates that involves all global qubits (i.e., $s = \#global\_qubits$) so that MPI communication is minimized. The depth (i.e., the number of layers) of the model circuit is set to 10. \reffig{fig:comparison_qiskitaer} plots the execution time of the QV model circuit with mpiQulacs running on Todoroki and Qiskit Aer running on the GPU cluster system. The results of Qiskit Aer is obtained from \cite{qiskitaer:2020}. 

\reffig{subfig:weak_scaling_QV_time} shows the weak scaling results of both the simulators. We here set the number of qubits to 30 per node and increment it with doubling the number of nodes. Qiskit Aer outperforms mpiQulacs when the 30-qubit circuit is executed on a single node. This is because Qiskit Aer uses six GPUs while mpiQulacs uses only one CPU. On the other hand, mpiQulacs significantly outperforms Qiskit Aer with more than 31 qubits. mpiQulacs achieves a high weak scalability to the number of qubits by reducing MPI communication with the fused swap gates. In contrast, Qiskit Aer increases the execution time linearly as the number of qubits is increased.

\reffig{subfig:strong_scaling_QV_time} shows the strong scaling results, where the number of nodes is varied from 1 to 64 with fixed 30 qubits. Note that Qiskit Aer uses six GPUs for the single-node simulation but only one GPU per node for the multi-node simulation. This graph shows that mpiQulacs outperforms Qiskit Aer when the numbers of A64FX CPUs and V100 GPUs are same (i.e., on two or more nodes). In particular, mpiQulacs achieves over 2x higher performance when the number of node is 16.

\begin{figure}[t]
  \centering
  \includegraphics[width=0.48\textwidth]{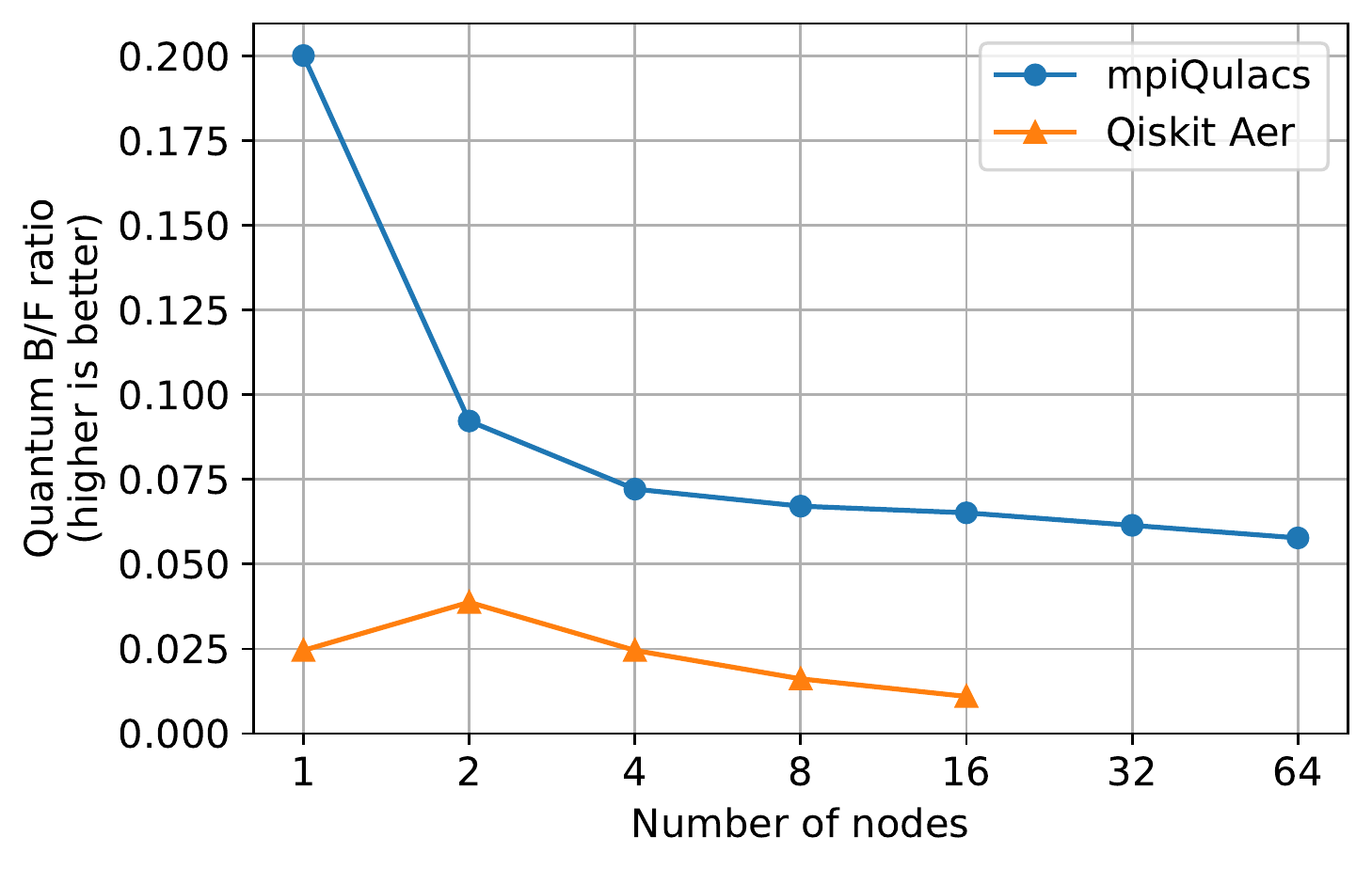}
  \caption{Quantum B/F ratio based on the strong scaling results of mpiQulacs running on Todoroki and Qiskit Aer running on the IBM GPU cluster system.}
  \label{fig:strong_scaling_QV_qbf}
\end{figure}

\reffig{fig:strong_scaling_QV_qbf} plots the QBF of both the simulators based on the strong scaling results shown in \reffig{subfig:strong_scaling_QV_time}. mpiQulacs achieves higher QBF than Qiskit Aer consistently across the varied number of nodes. The difference between them is especially significant (8x) when the number of node is one, because a single A64FX CPU (3.1 TFLOPS) achieves the 60\% performance of six GPUs (42.0 TFLOPS in total).

{\it Summary of comparison with Qiskit Aer}: mpiQulacs achieves a high weak/strong scalability by minimizing  MPI communication and outperforms Qiskit Aer significantly for large-scale simulation on multiple nodes. The QBF of mpiQulacs running on Todoroki is much higher than that of Qiskit Aer running on the GPU cluster system.

\subsection{Evaluation using Quantum Software Benchmark}

Finally, we evaluate three existing open-source distributed simulators (Intel-QS, QuEST, and Qiskit Aer) by ourself on the {\it AI Bridging Cloud Infrastructure (ABCI)} supercomputer and compare the results with that of mpiQulacs evaluated on Todoroki. We run QuEST only with CPUs because it does not support the distributed simulation using multiple GPUs. Unfortunately, we cannot evaluate JUQCS-G by ourself because it is not an open-source simulator.

\begin{table}[ht]
  \caption{Configuration of ABCI A-nodes}
  \begin{center}
  \begin{tabular}{l|l}
      \hline
      Cluster node & PRIMERGY GX2570M6 \\
      \# of nodes & 120 \\
      \hline \hline
      CPU & Xeon\textregistered\ Platinum 8360Y\\
      \# of CPUs per node & 2 \\
      \# of cores per CPU & 36 \\
      CPU memory & 512~GB DDR4 DRAM\\
      Theoretical peak FLOPS & 2.4 TFLOPS \\
      Theoretical peak memory bandwidth & 200 GB/s \\
      \hline
      GPU & NVIDIA A100 \\
      \# of GPUs per node & 8 \\      
      GPU memory & 40~GB HBM2 \\
      Theoretical peak FLOPS & 19.5 TFLOPS \\
      Theoretical peak memory bandwidth & 1,555 GB/s \\
      GPU interconnect & NVLink\textregistered\ 3 \\
      \hline
      Interconnect & InfiniBand HDR \\
      \# of HCAs & 4 \\
      \hline
      OS & RHEL 8.2 \\
      Compiler & GCC 9.3.0 \\
      CUDA Toolkit & CUDA 11.2.2 \\
      MPI & OpenMPI 4.0.5 \\
      \hline
  \end{tabular}
  \label{tab:abci_anode}
  \end{center}
\end{table}

 ABCI is an open supercomputer operated by National Institute of Advanced Industrial Science and Technology (AIST). Although two types of computing nodes called {\it V-node} and {\it A-node} are available on it, we only use the A-nodes because they have more high-end hardware resources than the V-nodes. \reftab{tab:abci_anode} summarizes the configuration of the A-nodes. There are 120 A-nodes, each of which has two sockets of Xeon\textregistered\ Platinum 8360Y processors with 512~GB DDR4 DRAM and eight NVIDIA A100 GPUs. The specification of the A100 GPU is identical to that equipped on JUWELS Booster, except the GPU counts per node. We calculate the theoretical peak FLOPS of each CPU by subtracting the total FLOPS of eight A100 GPUs per node from the total FLOPS of each node and dividing the difference by the number of CPUs per node (= (160.8 TFLOPS - 19.5 TFLOPS $\times$ 8) / 2) \cite{abci_user_guide}. In addition, we calculate the theoretical peak bandwidth of DDR4 DRAM based on the memory type and the number of memory channels supported by the CPU (= 3200 MHz $\times$ 8 bytes $\times$ 8 channels).

\begin{figure}[t]
  \centering
  \subfloat[Weak scaling (30 qubits per CPU, 31 qubits per GPU)]
  {\includegraphics[width=0.48\textwidth]{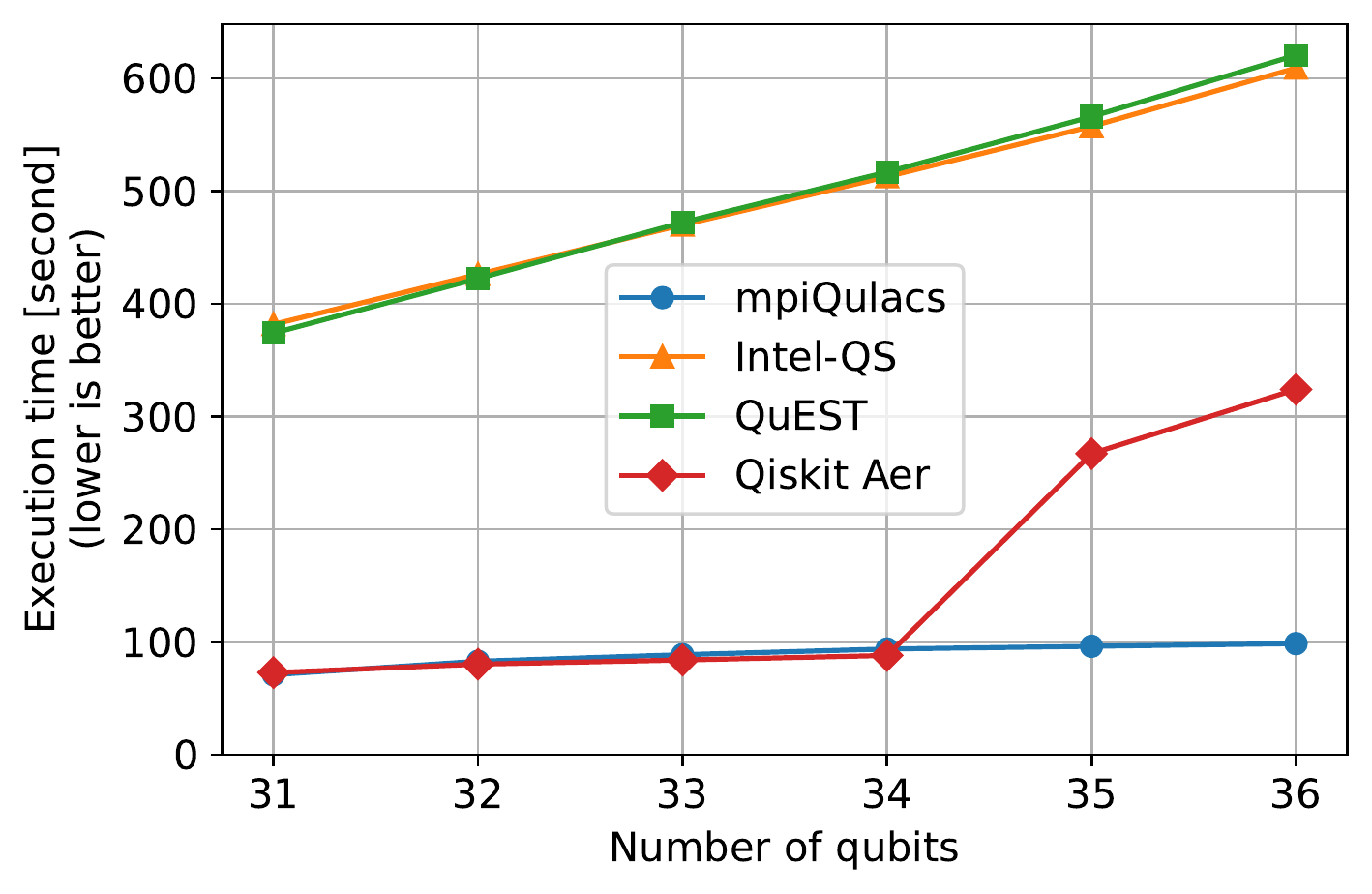}%
  \label{subfig:weak_scaling_QSB_time}}
  \\
  \subfloat[Strong scaling (fixed 30 qubits)]
  {\includegraphics[width=0.48\textwidth]{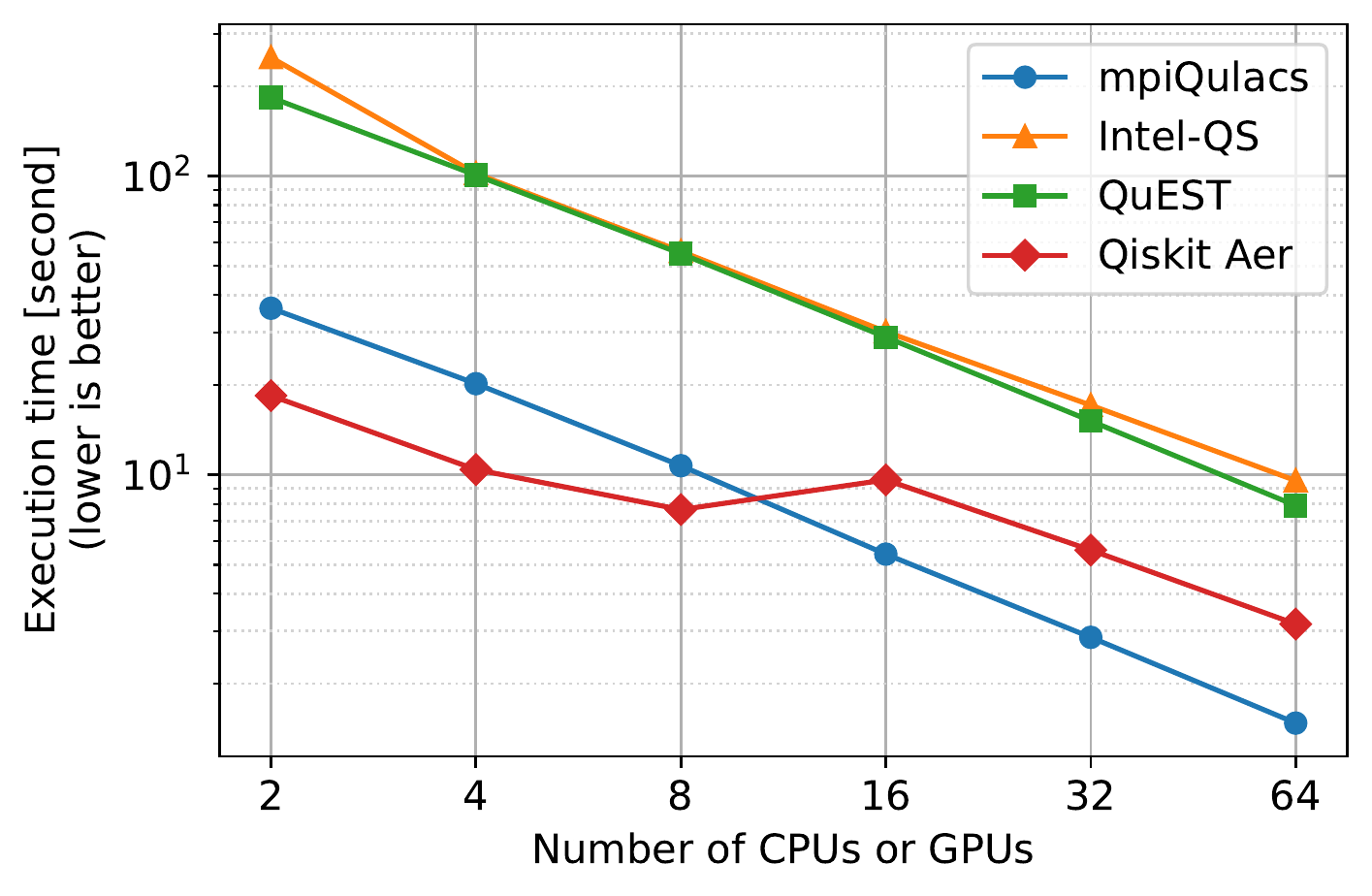}%
  \label{subfig:strong_scaling_QSB_time}}
  \caption{The execution time of Quantum Software Benchmark circuit with mpiQulacs running on Todoroki and Intel-QS, QuEST, and Qiskit Aer running on ABCI.}
  \label{fig:comparison_qsb}
\end{figure}

To compare mpiQulacs with the three existing simulators, we use the {\it Quantum Software Benchmark (QSB)} circuit \cite{qsb}, which is reviewed by the developers of representative simulators such as Qiskit and Cirq and used for the evaluation of the original version of Qulacs \cite{qulacs:2021}. This circuit consists of ten sets of a {\it rotation layer} and a {\it CNOT layer}, followed by one more rotation layer. In a rotation layer, random RZ, RX, and RZ gates are applied on each qubit. In a CNOT layer, CNOT gates are applied on the $i$-th target qubit and the $(i + 1\ mod\ n)$-th control qubit for $(0 \leq i < n)$ where $n$ is the number of qubits. For mpiQulacs, we insert fused-swap gates that involves all global qubits (i.e., $s = \#global\_qubits$) in the circuits so that the amount of MPI communication is minimized and the final state vector is consistent to the original circuit. \reffig{fig:comparison_qsb} plots the execution time of the QSB circuit with mpiQulacs running on Todoroki and Intel-QS, QuEST, and Qiskit Aer running on the ABCI A-nodes. 

\reffig{subfig:weak_scaling_QSB_time} shows the weak scaling results of the four simulators. We set the number of qubits to 30 per CPU and 31 per GPU and increment it with doubling the number of CPUs or GPUs. The performance of Intel-QS and QuEST is comparable because they use only CPUs. They increase the execution time  linearly along with the number of qubits due to the increase in the amount of MPI communication. In contrast, Qiskit Aer that is accelerated with GPUs significantly outperforms Intel-QS and QuEST. It achieves nearly ideal weak scaling with up to 34 qubits because eight GPUs on a single node are fully utilized. However, it steeply increases the execution time with 35 and 36 qubits due to inter-node MPI communication. In these cases, we run Qiskit Aer on 16 nodes and 32 nodes with a single GPU per node because the multi-node simulation with multiple GPUs per node is much slower. Interestingly, mpiQulacs achieves the comparable performance to Qiskit Aer with up to 34 qubits and sustains nearly ideal weak scaling with 35 qubits or more. The vectorization for A64FX CPUs and fused-swap gates of mpiQulacs contribute to this result.

\reffig{subfig:strong_scaling_QSB_time} shows the strong scaling results of the four simulators with the log-scaled y-axis. We here fix the number of qubits to 30 and vary the numbers of CPUs and GPUs from 2 to 64. mpiQulacs, Intel-QS, and QuEST achieve nearly ideal strong scaling, and mpiQulacs outperforms Intel-QS and QuEST significantly. Qiskit Aer outperforms mpiQulacs with up to eight GPUs, because Qiskit Aer uses multiple GPUs on a single node while mpiQulacs uses multiple nodes. In contrast, mpiQulacs outperforms Qiskit Aer when inter-node MPI communication is required. 

\begin{figure}[t]
  \centering
  \includegraphics[width=0.48\textwidth]{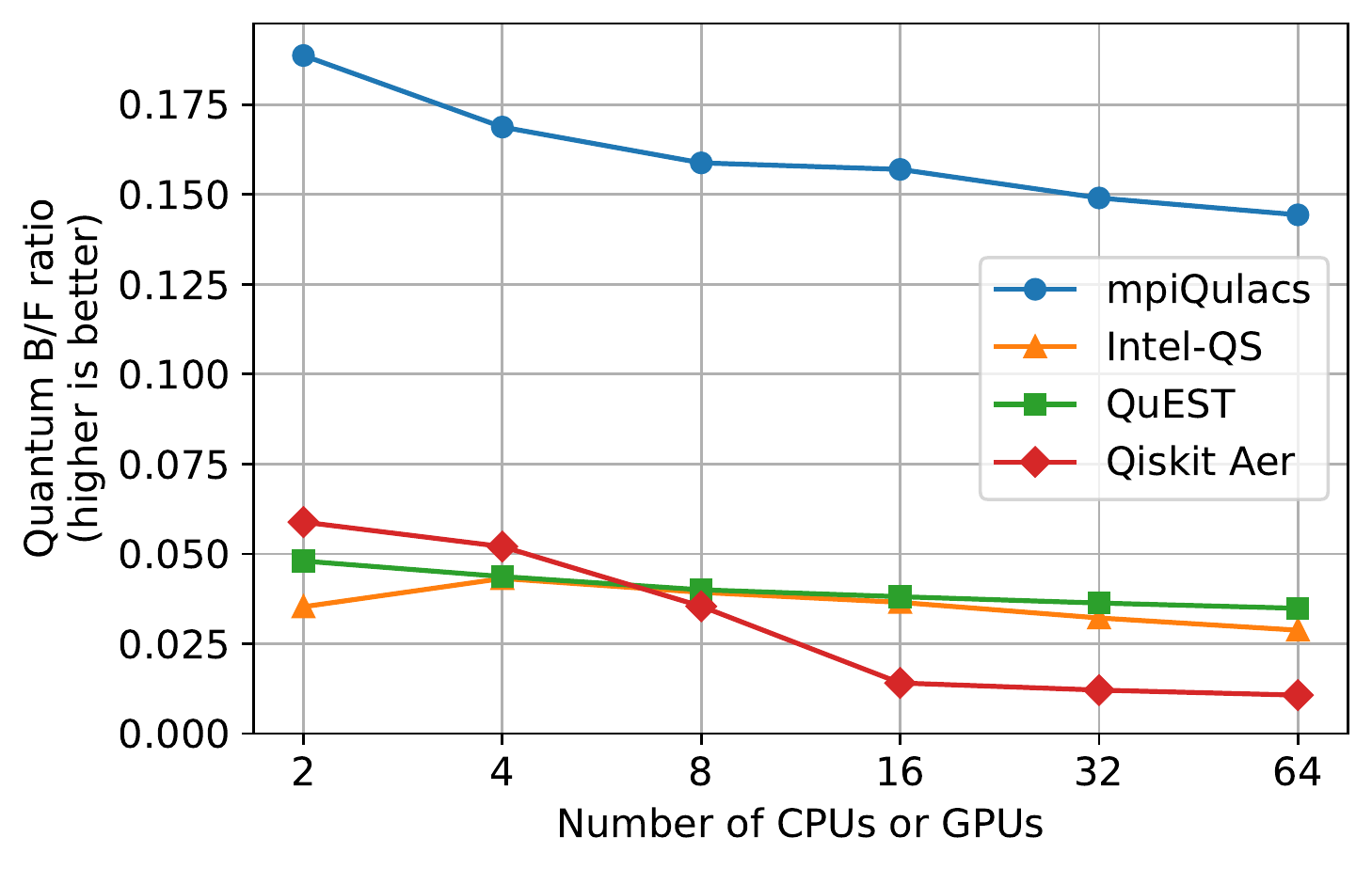}
  \caption{Quantum B/F ratio based on the strong scaling results of mpiQulacs running on Todoroki and  Intel-QS, QuEST, and Qiskit Aer running on ABCI.}
  \label{fig:strong_scaling_QSB_qbf}
\end{figure}

\reffig{fig:strong_scaling_QSB_qbf} shows the QBF of the four simulators based on the strong scaling results shown in \reffig{subfig:strong_scaling_QSB_time}. Interestingly, the QBF of GPU-accelerated Qiskit Aer is comparable or lower compared to that of CPU-based Intel-QS and QuEST. This is because the total FLOPS of GPUs used by Qiskit Aer is much higher than that of CPUs used by Intel-QS and QuEST, although Qiskit Aer significantly outperforms Intel-QS and QuEST as shown in \reffig{subfig:strong_scaling_QSB_time}. In contrast, mpiQulacs achieves much higher QBF than the other three simulators owing to the comparable performance to Qiskit-Aer and the comparable total FLOPS to Intel-QS and QuEST. This result means that mpiQulacs running on A64FX-based Todoroki fits the requirements of distributed state vector simulation.

\begin{figure}[t]
  \centering
  \includegraphics[width=0.48\textwidth]{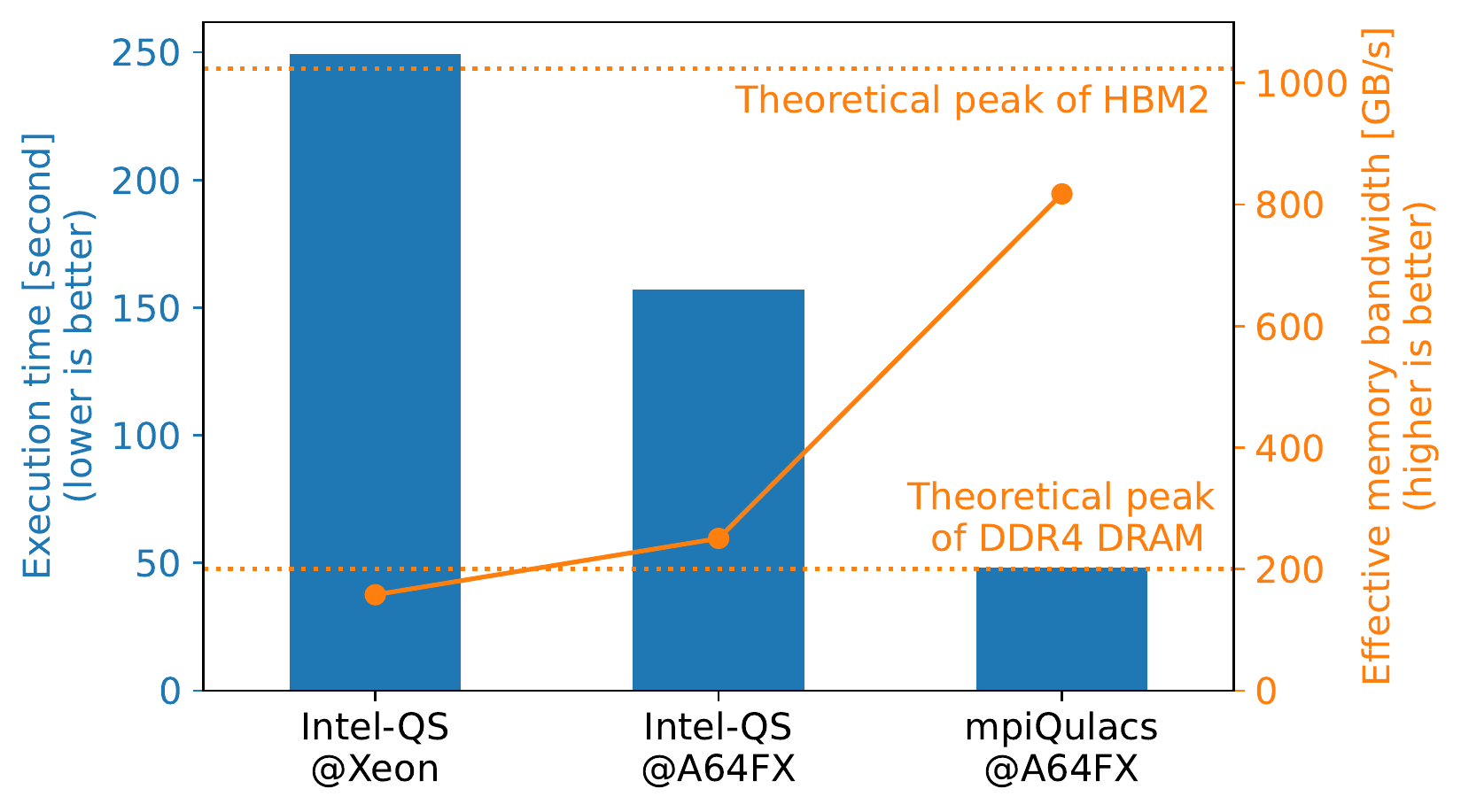}
  \caption{The execution time of the 30-qubit Quantum Software Benchmark circuit and effective memory bandwidth.}
  \label{fig:QSB_30qubits_1CPU}
\end{figure}

To analyze the benefit of the vectorization for the A64FX CPU of mpiQulacs, we execute the 30-qubit QSB circuit in three ways: Intel-QS with a single Xeon CPU on ABCI, Intel-QS with a single A64FX CPU on Todoroki, and mpiQulacs with a single A64FX CPU on Todoroki. \reffig{fig:QSB_30qubits_1CPU} shows the execution time of the QSB circuit on the left y-axis and the {\it effective} memory bandwidth on the right y-axis. The effective memory bandwidth is calculated by dividing the total amount
of memory traffic in bytes (= $2^{30+5} \times \#gates$) by the execution time. We can see in this figure that the performance of Intel-QS with the Xeon CPU is limited by the bandwidth of DDR4 DRAM. Thus, Intel-QS with the A64FX CPU reduces the execution time by 37\% by utilizing the high bandwidth of HBM2. However, in this case, the effective memory bandwidth is far from the theoretical peak bandwidth of HBM2. In contrast, mpiQulacs with the A64FX CPU achieves over 80\% of the theoretical peak bandwidth of HBM2 and thus further reduces the execution time by 69\%. This result demonstrates that mpiQulacs brings out the potential of the A64FX CPU.

{\it Summary of evaluation using QSB}: mpiQulacs running on Todoroki achieves nearly ideal weak/strong scaling and significantly outperforms Intel-QS, QuEST, and Qiskit Aer running on ABCI for large-scale simulation on multiple nodes. The QBF evaluation shows that mpiQulacs running on Todoroki fits the requirements of distributed state vector simulation. Moreover, we demonstrate that mpiQulacs fully utilizes the high memory bandwidth of the A64FX CPU.

\section{Conclusion and future work}
\label{sec:conclusion}

We develop mpiQulacs that is a fast and scalable distributed state vector simulator to accelerate the development of emerging quantum applications. It is optimized to fully utilize the high memory bandwidth of A64FX CPUs and supports a fused-swap gate to minimize the amount of MPI communication. We evaluate weak and strong scaling of mpiQulacs on the 64-node A64FX-based Todoroki cluster system using several quantum benchmark circuits. By comparing mpiQulacs with existing distributed state vector simulators, we show that mpiQulacs outperforms them for large-scale simulation on tens of nodes and achieves nearly ideal weak and strong scaling. Moreover, the evaluation in terms of quantum B/F ratio (QBF) demonstrates that mpiQulacs running on Todoroki fits the requirements of distributed state vector simulation. 

We are currently constructing another 1024-node A64FX-based cluster system, where mpiQulacs can simulate up to 40-qubit quantum circuits. We will use it to conduct research on quantum applications and further accelerate the development of real quantum applications. The nearly ideal scalability of mpiQulacs to larger-scale simulation will be more remarkable on this system. We also believe that the high QBF of mpiQulacs will contribute to lower energy consumption and/or lower system cost for such large-scale state vector simulation.

\section*{Acknowledgement}

The authors thank all of our colleagues who constructed the Todoroki cluster system. Their great effort helped us obtain the remarkable evaluation results of mpiQulacs.

\bibliographystyle{IEEEtran}
\balance
\bibliography{reference}

\end{document}